# Nanostructured Pt-Doped 2D MoSe$_2$: An Efficient Bifunctional Electrocatalyst for both Hydrogen Evolution and Oxygen Reduction Reactions

## Shrish Nath Upadhyay[1] and Srimanta Pakhira[1, 2, 3*]

[1] Department of Metallurgy Engineering and Materials Science (MEMS), Indian Institute of Technology Indore (IIT Indore), Khandwa Road, Simrol, Indore-453552, MP, India.
[2] Department of Physics, Indian Institute of Technology Indore, Khandwa Road, Simrol, Indore-453552, MP, India.
[3] Centre for Advanced Electronics (CAE), Indian Institute of Technology Indore, Khandwa Road, Simrol, Indore-453552, MP, India.
*Corresponding author: spakhira@iiti.ac.in (or) spakhirafsu@gmail.com

**Abstract:** Two-dimensional transition metal dichalcogenides (TMDs) are a new family of 2D materials with features that make them appealing for potential applications in nanomaterials science and engineering. Recently, these 2D TMDs have attracted a great research interest because of the abundant choice of materials with diverse and tunable electronic, optical, chemical, and electrocatalytic properties. Although, the edges of the 2D TMDs show excellent electrocatalytic performance, their basal plane (001) is inert which hinders the industrial applications for electrocatalysis. Transition metal/chalcogen atom vacancies or doping with some other foreign atoms may be a remedy to activate the inert basal plane. Here, we have computationally designed the 2D monolayer MoSe$_2$ and studied its electronic properties with electrocatalytic activities. Pt-atom has been doped in the pristine 2D MoSe$_2$ (i.e., Pt-MoSe$_2$) to activate the inert basal plane resulting zero bandgap. This study reveals that the Pt-MoSe$_2$ is an excellent bifunctional electrocatalyst for both the hydrogen evolution reaction (HER) and oxygen reduction reaction (ORR) with the aid of the first principle-based hybrid density functional theory (DFT). Periodic hybrid DFT method has been applied to compute the electronic properties of both the pristine MoSe$_2$ and Pt-MoSe$_2$. To determine both the HER and ORR mechanisms on the surface of the Pt-MoSe$_2$ material, a non-periodic DFT calculation has been performed by considering a molecular Pt$_1$-Mo$_9$Se$_{21}$ cluster model. The present study shows that the 2D Pt-MoSe$_2$ follows Volmer-Heyrovsky mechanism for HER with the energy barriers about 9.29 kcal.mol$^{-1}$ and 10.55 kcal.mol$^{-1}$ during the H$^{\bullet}$- migration and Heyrovsky reactions. The ORR is achieved by four-electron transfer mechanism with the formation of two transition energy barriers about 14.94 kcal.mol$^{-1}$ and 11.10 kcal.mol$^{-1}$, respectively. The lower energy barriers and high turnover frequency during the reactions expose that the Pt-MoSe$_2$ can be adopted as an efficient bifunctional electrocatalyst for both the HER and ORR. The present studies demonstrate that the exceptional HER and ORR activity and stability performance shown by the MoSe$_2$ electrocatalyst can be enhanced by Pt-doping, opening a promising concept for the sensible design of high-performance catalyst for H$_2$ production and O$_2$ reduction.

**1. Introduction**

Growing global energy crisis and environmental pollution are the major problems of the society these days which rise due to the large exploitation of conventional energy sources such as coal, natural gas and petroleum.[1] These primary problems have stimulated the interest of researchers in discovering and developing sustainable green energy storage and conversion systems.[2] Fuel cells, electrochemical water splitting and metal-air batteries can be the promising solution in this regard due to their high energy density and environmentally friendly nature.[3–5] Oxygen reduction reaction (ORR: O$_2$ + 4H$^+$ + 4e$^-$ = 2H$_2$O)[6] is the important cathodic reaction occurring in the fuel cells and its kinetics is comparatively slower than that of the reaction occurring at the anode side of the fuel cell.[7] Hydrogen evolution reaction (HER: 2H$^+$ + 2e$^-$ = H$_2$)[8,9] also occurs at cathode in the electrochemical water splitting which results in the production of hydrogen. An efficient electrocatalyst is required to increase the reaction rates of both HER and ORR in water splitting and fuel cells, respectively.[10,11] A few factors are required to keep in mind while selecting an efficient electrocatalyst for both the HER and the ORR, such as they must be cheap, earth-abundant and highly efficient towards the electrocatalytic activities.[11–14]

Ruthenium (Ru), Platinum (Pt) and Palladium (Pd) are a few noble metals which have shown the best electrocatalytic activity in fuel cells for ORR and water splitting for HER.[15,16] The Pt electrode has nearly zero over potential in acidic electrolytes, and that is why it has shown the best electrocatalytic activity till to date. Though Pt electrode is the best electrocatalyst, its commercial-scale application is impossible due to the metal scarcity and high cost.[17,18] In other words, Pt prohibits its application to fulfill the energy demands and commercialization industrial applications. It is thus a challenging task to find out an earth-abundant, cheap, and highly efficient electrocatalyst which can be used as an electrode for the HER in electrochemical water splitting to produce $H_2$ and for the ORR in fuel cells for the conversion of chemical energy into electrical energy at the commercial or industrial scale.

In the last few decades, 2D materials have attracted a lot of attention to the researchers working in the field of nanomaterial science, materials physics and engineering. The family of these 2D materials consists of two-dimensional crystalline materials with thickness varying from one atomic layer to more than ten nanometers. They have a wide spectrum of electronic and optical properties covering metals, semimetals, semiconductors with various energy band gaps and insulators. These 2D materials not only have wide range of electronic properties but also have dangling bonds, high surface areas and surface state free nature, and they may exhibit strong spin−orbit coupling and quantum spin Hall effects which enable them to be studied in various applications.[19,20] Among the various 2D materials, Transition Metal Dichalcogenides (TMDs) have been extensively studied for electrocatalytic applications such as ORR and HER. The reason behind the efficient electrocatalytic activity of these 2D TMDs is that all of their active edge sites are exposed in the reactions due to the atomic thin nature of their nanosheets.[21,22] Layered TMDs consist of sheets of transition metal atoms in an alternating manner which is sandwiched in between two layers of chalcogen atoms.[23] Due to the difference in oxidation states of transition metals (TMs) and chalcogen atoms, they are bonded to each other with the strong ionic bonds, which provide strength to the nanosheets. At the same time, there exists a weak van der Waals (vdW) interaction between two successive layers which favors the exfoliation of the nanosheets.[9,24,25] A variety of combinations can be realized for the transition metals and chalcogen atoms which opens a broad platform for studying the electrocatalytic activity for HER and ORR reactions.

The exposed edges of the 2D TMDs are highly active towards the electrocatalytic applications. However, the overall electrocatalytic performance of these 2D TMDs is limited due to several constraints. Basal planes (001) of the TMDs are usually catalytically inert and hence they cannot be used for the efficient electrocatalysts in the industrial applications as well as commercialization purposes. The key point here is to activate the basal plane of the 2D monolayer TMDs so that it can be further used in electrocatalytic reactions as an efficient electrocatalyst.[26] Many efforts have been made in this context to activate the inert basal planes of the TMDs, such as creating some defective sites in the lattice or doping it with some foreign atoms.[23] Noble metal-supported TMDs have been explored with efficient electrocatalytic activity towards ORR and HER.[27,28] But commercialization of these noble metal supported TMD electrocatalysts is difficult because of the cost factors and scarcity of the noble metals. In the family of 2D TMDs, molybdenum diselenide ($MoSe_2$) is being highly explored after molybdenum disulfide ($MoS_2$). Like other 2D TMDs, $MoSe_2$ has also tunable electronic band gap which makes it as a promising candidate for various applications. The layered structure of $MoSe_2$ plus the size and electrical conductivity of Se provide a good opportunity for hosting counter ions in electrochemical energy storage systems such as lithium-ion and sodium-ion batteries.

In this manuscript, we have computationally designed a 2D monolayer pristine $MoSe_2$ material and then it has been extended to a 2x2 supercell to dope Pt atom in the pristine $MoSe_2$ to form 2D monolayer Pt-$MoSe_2$ material. The first principle-based periodic hybrid density functional theory (i.e., the DFT-D method) was used to obtain the equilibrium 2D monolayer structures of both the pristine $MoSe_2$ and Pt-doped $MoSe_2$ materials and the electronic properties calculations i.e., band structure and the total density of states (DOS). From the electronic property calculations, it was found that the 2D single layer $MoSe_2$ material has a higher energy bandgap about 2.21 eV and no electron density was present at the Fermi energy ($E_F$) level. After Pt doping, the energy band gap of the 2D monolayer Pt-$MoSe_2$ material has been reduced to 0 eV with the enough electronic density around the Fermi energy ($E_F$) level, which indicates that the 2D Pt-$MoSe_2$ may be a promising bifunctional electrocatalyst for both the ORR and HER. We developed a non-periodic molecular cluster model system of $Mo_{10}Se_{21}$ to explore the reaction paths and to compute intermediates as well as the reaction barriers for HER on the active surface of 2D monolayer $MoSe_2$. Similarly, a non-periodic molecular cluster model system of $Pt_1$-$Mo_9Se_{21}$ was developed for the 2D monolayer Pt-$MoSe_2$ material as shown in Fig. 1. This 2D monolayer cluster model system of the 2D Pt-$MoSe_2$ (i.e., $Pt_1$-$Mo_9Se_{21}$) material was used for exploring the reaction paths, reaction kinetics, and reaction barrier energies of both the HER and ORR using first the principle-based DFT methods.

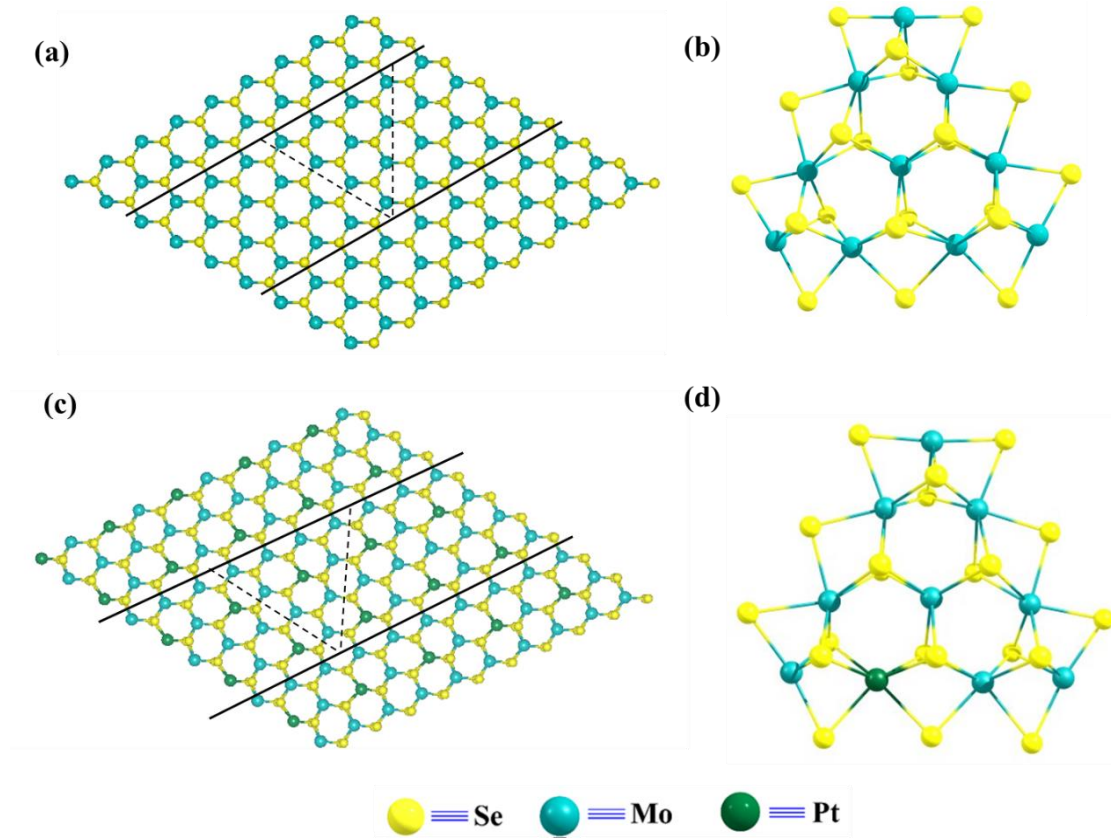

**Fig. 1.** Top views of (a) periodic equilibrium structure of 2D monolayer MoSe$_2$, the inside dotted triangle represents the non-periodic molecular cluster model. The two horizontal solid lines reveal the ends along the ($\bar{1}010$) Se-edge and ($10\bar{1}0$) Mo-edge of the 2D monolayer MoSe$_2$ material; (b) Mo$_{10}$Se$_{21}$ finite non-periodic molecular cluster model for 2D monolayer MoSe$_2$; (c) periodic equilibrium structure of 2D monolayer Pt-MoSe$_2$, the inside dotted triangle represents the non-periodic molecular cluster model. The two horizontal solid lines reveal the ends along the ($\bar{1}010$) Se-edge and ($10\bar{1}0$) Pt-/Mo-edge of the 2D monolayer Pt-MoSe$_2$ material; (d) Pt$_1$-Mo$_9$Se$_{21}$ finite non-periodic molecular cluster model for 2D monolayer Pt-MoSe$_2$, are shown here.

## 2. Methods and computational details

For the periodic 2D slab structure computations, 2D single layer of the pristine MoSe$_2$ and Pt-MoSe$_2$ were terminated at the Mo-/Pt-edges ($10\bar{1}0$) and Se-edges ($\bar{1}010$) boundaries as depicted in Fig. 1a-d. First, we performed our study on the 2D pristine MoSe$_2$ to obtain the equilibrium geometry. Then, we extended the single unit cell of the 2D monolayer MoSe$_2$ to 2x2 supercell to dope the Pt atom to create monolayer 2D Pt-doped MoSe$_2$ i.e., Pt-MoSe$_2$. Out of total four Mo atoms of the 2x2 supercell of MoSe$_2$, we replaced one Mo atom by one Pt atom to obtain the 2D monolayer Pt-MoSe$_2$ material and to study the electronic properties of this material. Equilibrium geometry of the 2D MoSe$_2$ was found to have the equilibrium lattice parameters $a = b = 3.234$ Å with the "***P-6m₂***" space group symmetry and the equilibrium lattice constants of the 2D monolayer Pt-MoSe₂ material were $a = b = 6.738$ Å with the "***P1***" space-group symmetry.

**2.1 Periodic DFT calculations.** All the geometries, 2D layer structures and electronic properties calculations of both the 2D monolayer pristine MoSe₂ and Pt-MoSe₂ have been performed by using first-principles based B3LYP-D3 dispersion-corrected density functional theory (DFT-D) method[29–31] implemented in *ab initio-based* CRYSTAL17 suit code[29,32,33]. CRYSTAL17 program uses Gaussian type basis sets (GTO i.e., Gaussian type orbitals)[34] for all the atoms during the computation. This GTO basis sets are more effective than plane wave basis sets for the hybrid DFT calculations. In the present calculations, triple zeta polarization quality basis sets have been used for the O, H, Mo and Se atoms, and similarly Pt_doll_2004 basis set with the effective core potentials (ECPs) has been used for the Pt atom.[35,36] In this computational work, we have included the semi-empirical Grimme's 3$^{rd}$ order (-D3) dispersion corrections to explain the weak van der Waals (vdW) interactions between different layers of MoSe₂ as well as Pt-MoSe₂ and among various atoms.[37–41] The DFT-D method is very useful tool to obtain fine geometries, energy and density as these are less affected by spins in this computational method.[40,42–47] To consider the electrostatic potential in these calculations, we have developed vacuum slabs and the energies have been reported with respect to vacuum (w.r.t. vac.) by contemplating electrostatic potential and its derivatives. The electrostatic potential due to the crystal/2D layer atoms net charge is present in almost all 2D slabs, except pure covalent ones. The electrostatic potential is described by the series which is divergent for an infinite crystal. Calculation of electrostatic potential between different atomic configurations is necessary for the first principles based DFT calculations. The exact electrostatic potential V, its derivatives E (electric field) and E′ (Electric field gradient) are evaluated at the same level of theory. The calculations of V (z), E(z), E′(z) and ρ(z) averaged in the volume between z–ZD and z+ZD (2D only) are done by setting the ICA value '3' and NPU value '0' (these properties are computed at the atomic positions) for 2D systems implemented in CRYSTAL17 program.[29, 32-34]

After obtaining the equilibrium geometries (using Monkhorst[48] k-mesh grids of 4x4x1) of both the pristine 2D monolayer MoSe₂ and 2D monolayer Pt-MoSe₂ materials, we performed the electronic properties i.e., the electronic band structures and the total density of state (DOS) calculations at the same level of theory with respect to the vacuum. All the integrations in the first Brillouin zone were sampled on 20x20x1 Monkhorst-pack[48] k-mesh grids for the pristine 2D monolayer MoSe$_2$ and Pt-MoSe$_2$ materials. The threshold value was set to $10^{-7}$ a.u for the convergence of forces, energies, and electron densities and 0.01 Hartree value of smearing was used. The electronic band structure was plotted in the high symmetric k-direction, and the k-vector path was chosen as $\boldsymbol{\Gamma - M - K - \Gamma}$ in the first Brillouin zone w.r.t. vac. Total eight numbers of energy bands are computed around the Fermi energy level (E$_F$) to determine the electronic properties of the materials. The total density of states (DOS) as well as the *d*-subshell density of states contribution of the platinum (Pt) atom have been calculated at the equilibrium structures of both the materials studied here. To visualize all the equilibrium structures, VESTA visualization software is used.[49] All the equilibrium structures with their crystallographic information files (.cif) involved in the subject reaction have been provided in the Supplementary Information.

**2.2 Finite molecular cluster modelling and DFT calculations.** We have computationally developed a non-periodic finite molecular cluster model system Mo$_{10}$Se$_{21}$ of the 2D monolayer MoSe$_2$ as shown in Fig. 1b to explore the HER mechanism with the equilibrium geometries and reaction barriers. This Mo$_{10}$Se$_{21}$ cluster model system consists of 10 Mo atoms and 21 Se atoms to represent the parent 2D monolayer MoSe$_2$ periodic slab structure as shown in Fig. 1a. This model structure has completed the dangling bond by adding one extra Se atom in the system. Fig. 1 shows the termination scheme using two horizontal lines along the ($00\bar{1}0$) Mo-edges and Se-edges. The two triangles have been used to represent the terminations for Mo-edge and Se-edge clusters. We have considered the same triangle to set the dangling bonds in the finite cluster shown in Fig. 1 and one extra Se atom was considered in the molecular model system to neutralize the dangling bonds. The oxidation state of Mo atoms in the basal plane (001) of the finite molecular cluster model is +4. These Mo atoms are bonded with 3 Se atoms in upper and as well as in lower plane. This kind of geometry gives a contribution of 4/6 = 2/3 electrons towards each Mo-Se bonding resulting a stabilized structure. The oxidation state of Se is -2 and it is bonded with 3 Mo atoms which results a contribution of 2/3 electrons towards each Mo-Se bond. Similarly, the edges of the molecular cluster ($00\bar{1}0$) is being stabilized with the 2 local electrons Mo-Se bond having a single electron contribution towards 4 Mo-Se bonding in the basal plane. This 14/3 {i.e., (2×1) + [4× (2/3)]} contribution of electrons towards the Mo-Se bonding of the edge Mo atom is satisfied with the d$^2$ configuration of one Mo atom and d$^1$ configuration of two Mo atoms at the edges. The molecular system with this configuration has periodicity 3 which results the achievement of a stabilized molecular cluster model having three edges without any unsatisfied valency. Thus, we have considered the molecular cluster model $Mo_{10}Se_{21}$ for our $MoSe_2$ following the previous works on 2D monolayer Mn-doped $MoS_2$ and MoSSe Janus TMDs as shown in Fig. 1b.[50,51] This is a well-established molecular cluster model system to investigate the electrocatalytic activities such as HER, and there are several studies about the cluster model systems like $MoS_2$, MoSSe and Mn-doped $MoS_2$ which can be found in References 9, 50 and 51. The results showed that the Se-edges of the 2D $MoSe_2$ have similar hydrogen binding as the Mo-edge of $MoSe_2$, i.e., the two edges of $MoSe_2$ have approximately equal HER activity.[52] In the present study we have chosen the 50% Se coverage of Mo-edge to study the HER mechanism. Literature reviews have shown that 50% chalcogen coverage of Mo-edge is the most stable form.[52–55] Similarly, a non-periodic finite $Pt_1Mo_9Se_{21}$ molecular cluster model system has been considered to explore the HER and ORR mechanisms on the surfaces of the 2D monolayer Pt-doped $MoSe_2$ TMD as shown in Fig. 1c-d.

We have considered these molecular cluster models ($Mo_{10}Se_{21}$ and $Pt_1Mo_9Se_{21}$) of the 2D monolayer $MoSe_2$ to explore the HER mechanism. The molecular cluster model has been developed in such a way that it consists of similar properties compared to the periodic 2D monolayer $MoSe_2$. This molecular cluster model $Mo_{10}Se_{21}$ has been developed to investigate the HER mechanism with the most favorable reaction pathways as it allows more flexibility in the accuracy to use M06-L DFT method which is more precise to determine the reaction barriers, kinetics, and bond energies. It is easy to use cluster model with net charges which is difficult in the case of periodic system. Moreover, the cluster model allows us to incorporate the electrons ($e^-$) and protons ($H^+$) in various reaction steps and report free energies as a function of electrochemical potential and pH. The molecular cluster model has the same chemical properties compared with the periodic 2D monolayer $MoSe_2$. Another point, the cluster model system is big enough to represent the 2D monolayer structure of both the $MoSe_2$ and Pt-$MoSe_2$ which is computationally and quantum mechanically very costly. This model system is enough to explain the electrochemical activities of both the systems studied here. It should be noted here that the thermodynamic stability has been checked of the $Pt_1Mo_9Se_{21}$ system after that the Pt-doping in the pristine $Mo_{10}Se_{21}$ considering the finite cluster model systems and the present DFT computations reveal that the $Pt_1Mo_9Se_{21}$ is thermodynamically favorable by an amount of energy, $\Delta G$ = -20.597 eV and $\Delta H$ = -20.521 eV, with respect to the pristine one $Mo_{10}Se_{21}$.

The unit cell of the pristine $MoSe_2$ has been extended to 2x2 supercell and one Mo atom was replaced by one Pt atom to develop the 2D monolayer Pt-$MoSe_2$ periodic structure as shown in the Fig. 1c. After obtaining the equilibrium geometries, the structures of 2D pristine monolayer $MoSe_2$ and Pt-$MoSe_2$ are found in 2H phase. The equilibrium structures of the 2D monolayer pristine $MoSe_2$ and Pt-doped $MoSe_2$ TMDs are provided in the Supplementary Information in the form of crystallographic information file (.cif). Both the structures were optimized to locate the minimum energy (negative value of E) of the system which results in thermodynamic stability of these two equilibrium geometries. However, we have checked the thermodynamic stability of the system Pt-$MoSe_2$ after replacing one Mo atom by Pt atom and found that it is thermodynamically stable with negative value of $\Delta E$ = -23.57 eV. We have considered a $Pt_4$ cluster adsorbed on the surfaces of the 2D $MoSe_2$ (the ), and the present computations found that the structure is not stable computationally and the system showed convergence failure which indicates that it is not possible to get $Pt_4$ adsorbed structure on the surfaces of $MoSe_2$. Similarly, we have also computed the single Pt atom adsorbed on the surfaces of $MoSe_2$ and the values of $\Delta E$, $\Delta G$ and $\Delta H$ are about -3.56 eV, -3.23 eV and -3.55 eV, respectively, computed by the DFT method. This outcome indicates that the single Pt atom adsorption on the surfaces of the 2D monolayer $MoSe_2$ is less stable than the Pt-doped $MoSe_2$. This result tells that the doping of Pt atom in the pristine 2D monolayer $MoSe_2$ is more thermodynamically and energetically favorable compared to the Pt atom adsorption on the surfaces of it. Then after, one Mo atom in the $Mo_{10}Se_{21}$ finite molecular cluster system was replaced by one Pt atom to develop $Pt_1$-$Mo_9Se_{21}$ finite molecular cluster model for Pt-$MoSe_2$ as shown Fig. 1d. The M06L[56,57] DFT method was used to study the HER mechanism on the active surface of 2D monolayer pristine $MoSe_2$ taking account the $Mo_{10}Se_{21}$ finite non-periodic molecular cluster model system. Similarly, the same M06L[56,57] DFT method was employed to investigate both the ORR and HER mechanisms on the active surfaces of the 2D monolayer Pt-$MoSe_2$ considering the $Pt_1$-$Mo_9Se_{21}$ finite non-periodic molecular cluster model system. An earlier report says that the method DFT-M06L gives authentic energy barriers for reaction mechanisms of transition metal-based catalysts.[8] For all the computations, we have used 6-31++G* basis sets for the O, H, and Se atoms, and LANL2DZ basis sets for the Pt and Mo atoms with effective core potentials (ECPs).[58] We have used 6-31++G* Gaussian basis sets (Pople type of basis sets) for the O, H, and Se atoms, and LANL2DZ basis set for both the Mo and Pt atoms which utilizes Los Alamos effective core potential on the transition metal. In *ab initio* quantum chemistry calculations, Pople-type split valence basis sets are extensively used and give much validated results.[59] In the present study, we have used 6-31++G* double-zeta Pople-type basis set for the O, H, and Se atoms. For the metal atoms, LANL2DZ (Los Alamos National

Laboratory 2 double-zeta) has been employed in the present computations, which is a widely used effective core potential (ECP)-type basis set.[60] This mixed kind of basis sets have been extensively used to study the transition metal containing systems in density functional theory methods to explore the chemical reactions which provide reliable results.[50,51,59]

The Zero-Point Vibrational Energy (ZPE) and frequencies were calculated at the equilibrium geometry of all the intermediates using the same methods and basis sets. Transitions states (TSs) were observed and confirmed by obtaining only one imaginary frequency (negative value) in the modes of vibrations with intrinsic reaction coordinate (IRC) calculations.[42,45] All the computations for the ORR and HER mechanism were performed with the general-purpose electronic structure quantum chemistry program Gaussian16[61] to obtain the equilibrium geometries, TSs, energies, reaction barriers, etc.[61] All the equilibrium structures with their cartesian coordinates (finite molecular cluster model systems) involved in the subject reaction have been provided in the Supplementary Information.

It should be mentioned here that we have studied the HER mechanism on the surface of 2D monolayer $MoSe_2$ and Pt-$MoSe_2$ by taking into consideration the adsorption free energies of various reaction steps and calculating two TSs (Volmer and Heyrovsky). Using molecular clusters to model a periodic system for determining reaction mechanisms allows more flexibility in the accuracy of the methods. This allows us to use M06L which is more accurate for reaction barriers and bond energies than other computational methods (such as PBE, B3LYP, etc.),[7] the most common method for periodic calculations. We have focused on the electrocatalytic activities (HER and ORR) of the 2D monolayer $MoSe_2$ and Pt-$MoSe_2$ TMDs and the chemical reaction mechanism of both the HER and ORR in the present study. Thus, we have excluded the discussions about the potential dependence of the free energies and activation barriers of the electrochemical steps in the reaction mechanisms.

## 2.3 Validation of the cluster model

To validate that the cluster model has the same chemical properties as the periodic Mo-edge, we calculated the binding energy of a hydrogen atom to both the cluster and the periodic slab, both under vacuum conditions followed by previous works.[50,51] In both the cases, we reference the free H atom energy to that of 1/2 $H_2$ molecule. The hydrogen adsorption energy (electronic energy) has been calculated considering the finite non-periodic molecular cluster models of both the pristine $MoSe_2$ and Pt-$MoSe_2$ TMDs as well as their periodic 2D slab models to validate the chemical properties of both the cluster and periodic systems. The calculated values of hydrogen adsorption energies for both the systems considering both the model systems are almost equal as depicted in Fig. 2.

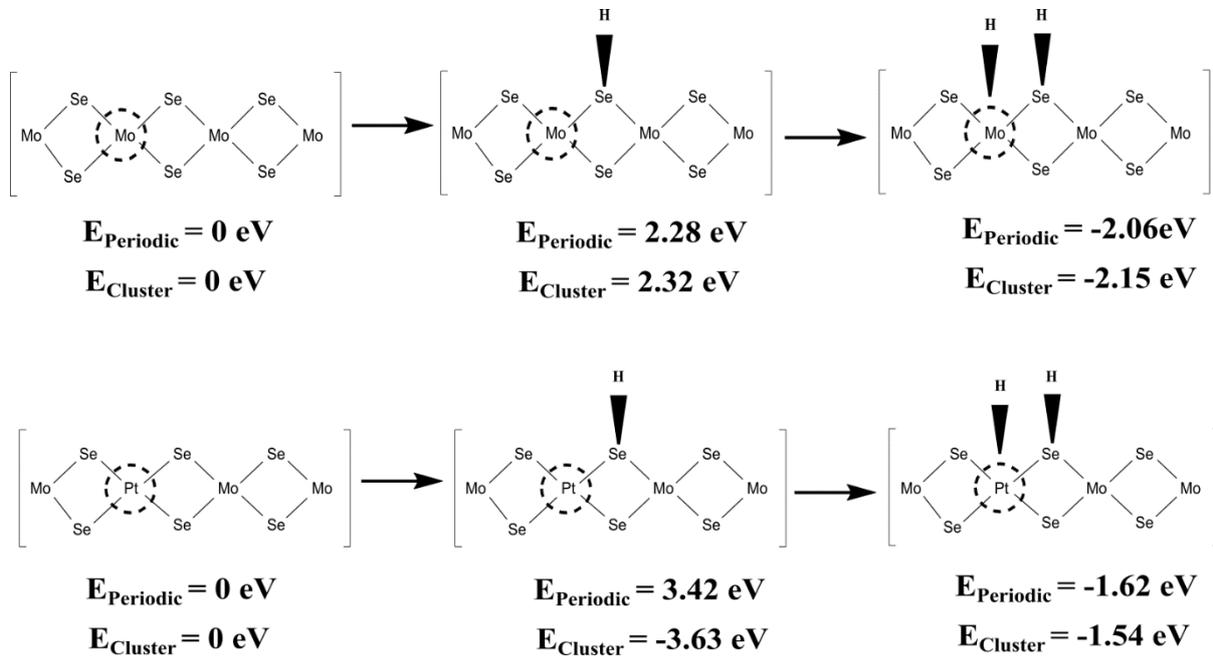

Fig. 2 Hydrogen adsorption energies on the surfaces of the cluster and periodic model systems of both the pristine 2D monolayer MoSe$_2$ and Pt-MoSe$_2$.

**2.4 HER and ORR mechanism.** The overall H$_2$ evolution reaction pathway is followed by Volmer-Heyrovsky or Volmer-Tafel mechanism irrespective of the medium which is given in the following equations below.

Volmer Reaction: $\quad * + H^+ + e^- \rightarrow H^*_{ads}$ (i)

Heyrovsky Reaction: $\quad H^*_{ads} + H^+ + e^- \rightarrow H_2 + *$ (ii)

Tafel Reaction: $\quad 2H^*_{ads} \rightarrow H_2 + *$ (iii)

The ORR reaction can occur either via dissociative or associative mechanisms. The dissociative and associative mechanism pathways are given by the equations (iv) and (v), respectively. The oxygen reduction reaction is $O_2 + 4H^+ + 4e^- \rightarrow 2H_2O$.

$O_2 \rightarrow O_2^* \rightarrow 2O^* \rightarrow O^* + 2OH^* \rightarrow 2O^* + H_2O$ (or $H_2O_2$) $\rightarrow OH^* + H^* \rightarrow H_2O$ (iv)

$O_2 \rightarrow O_2^* \rightarrow OOH^* \rightarrow O^* + H_2O$ (or $H_2O_2$) $\rightarrow OH^* + H_2 \rightarrow 2H_2O$ (v)

$H^+ + OH^- \longrightarrow H_2O$ (vi)

where * denotes an adsorption site or active site of the catalyst. Here, it should be noted that inclusion of the protons and electrons in equations (iv) and (v) have been left out for clarity, and H$_2$O$_2$ is the product of 2e$^-$ pathways of ORR.

**2.5 Theoretical calculations and equations.** The catalytic performances of both the 2D monolayer pristine MoSe$_2$ and Pt-MoSe$_2$ were characterized by the calculations of the changes of free energy (ΔG), enthalpy (ΔH), and electronic energy (ΔE) for all the reactant, products, intermediates, and TSs for both the HER and ORR.

Change of free energy:  $\Delta G = \sum G_{Product} - \sum G_{Reactant}$  (vii)

Change of enthalpy:  $\Delta H = \sum H_{Product} - \sum H_{Reactant}$  (viii)

Change of electronic energy:  $\Delta E = \sum E_{Product} - \sum E_{Reactant}$  (ix)

The Gibbs's free energy of an electron (e⁻) has been calculated at the standard hydrogen electrode (SHE) conditions where electron (e⁻) and proton (H⁺) (pH = 0) are in equilibrium with 1 atm H$_2$. To compute the solvation effects on the reaction barriers, polarizable continuum model (PCM) calculations have been performed by using water as a solvent with dielectric constant of 80.13. The PCM was employed for all the theoretical calculations to describe the solvation effects in the M06-L DFT computations, and three H$_2$O molecules with one H$_3$O⁺ (i.e., 3H$_2$O_H$_3$O⁺) have been combined unambiguously for the Heyrovsky reaction step. PCM is one of the best models to consider the solvation effects, and it is a commonly used method in computational chemistry to model solvation effects

**3. Results and discussion**

We have performed a theoretical study to obtain the equilibrium structures of both the pristine 2D monolayer MoSe$_2$ and 2D monolayer Pt-MoSe$_2$ materials and calculated the electronic properties (i.e., the electronic band structures and total density of states (DOS)) calculations for both the materials. The equilibrium 2D structure and electronic properties of the pristine 2D monolayer MoSe$_2$ computed by the DFT-D method are depicted in Fig. 3. Our present DFT-D study shows that the pristine 2D monolayer MoSe$_2$ has hexagonal ***P-6m$_2$*** symmetry with the equilibrium lattice constants *a* = *b* = 3.234 Å, α = β = 90º, and γ = 120º and the equilibrium lattice parameters along with the average Mo-Se bond length are reported in Table 1. The solid lines in the top view of Fig. 1a represents the one-unit cell of pristine 2D monolayer MoSe$_2$. The present calculations are in good agreement with the previously reported (*a* = *b* = 3.288 Å) experimental study performed by James and co-workers.[62] The electronic properties computations (i.e. the electronic band structures and total DOS) have been performed at the equilibrium structure of the 2D monolayer MoSe$_2$ material using the same DFT level of theory as displayed in Fig. 3b-c. The electronic band structures of the pristine 2D monolayer MoSe$_2$ have been plotted in the high symmetric direction. *Γ-M-K-Γ* w.r.t. vac. as depicted in Fig. 3b. The Fermi energy level (E$_F$) was found at -5.38 eV highlighted by dotted blue line as shown in Fig. 3b-c. The band structure calculations show that the pristine 2D MoSe$_2$ has a direct bandgap about 2.21 eV at the K-point as shown in Fig. 3b, which can also be observed from the DOS calculations as depicted in the Fig. 3c. The obtained energy band gap of 2D monolayer MoSe$_2$ in the present calculations is well harmonized with the earlier findings.[63] The computed electronic band gap of the 2D monolayer pristine MoSe$_2$ TMD reproduces the previous experimental and theoretical values within 0.03 eV.[63,64]

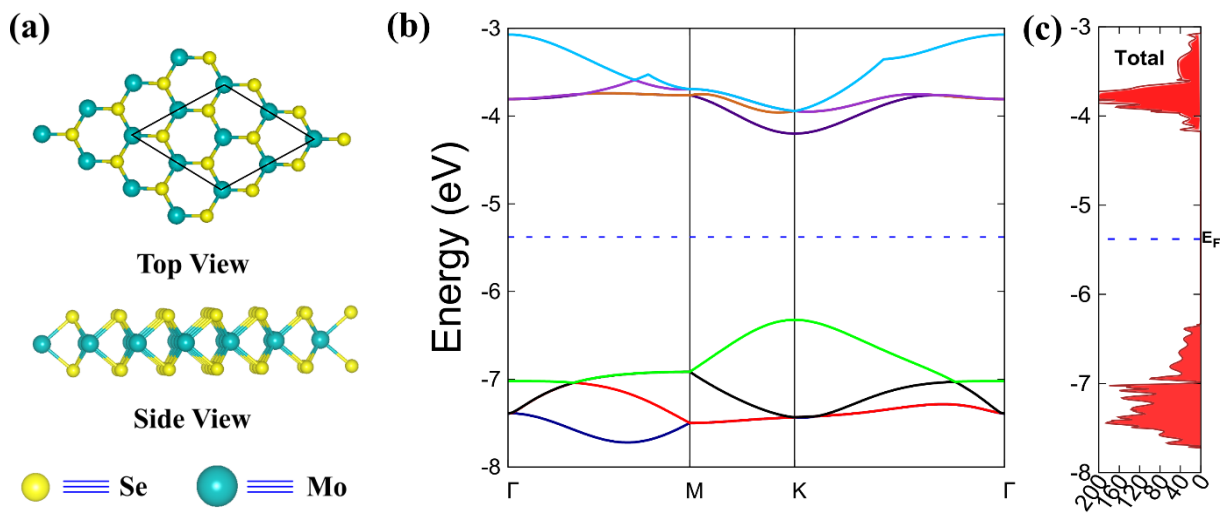

**Fig. 3.** (a) The top and side view of equilibrium structure of the pristine 2D monolayer MoSe$_2$; (b) electronic band structure, and (c) total density of states (DOS) of the pristine 2D monolayer MoSe$_2$ material computed at the DFT-D level of theory are shown here.

Introducing the Pt-doping in the pristine 2D single layer MoSe$_2$ material, the equilibrium structure and lattice parameters were significantly changed due to the Pt atoms in the unit cell. The elevated equilibrium geometry and lattice parameters of the 2D monolayer Pt-MoSe$_2$ material were calculated using the same DFT-D method and the values are *a = b =* 6.738 Å, α =β = 90º and γ = 121.7º with ***P1*** symmetry. The equilibrium average Mo-Se and Pt-Se bond lengths were 2.518 Å and 2.623 Å, respectively, and the Mo-Se bond length was elongated due to Pt-doping in the pristine MoSe$_2$. The structural parameters and lattice constants of the 2D monolayer Pt-MoSe$_2$ material are listed in the Table 1. To identify the phase transition of the 2D monolayer Pt-MoSe$_2$ from 2H to 1T phase, both the structures (2H-Pt-MoSe$_2$ and 1T-Pt-MoSe$_2$) were optimized to obtain the lattice parameters and formation energies. The phase transition can be identified by the energy difference ($\Delta E_{1T-2H}$) between 1T and 2H phases of the 2D monolayer Pt-MoSe$_2$ TMD and the energy difference has been calculated by using the following equation:

$$\Delta E_{1T-2H} = E_{1T} - E_{2H} \qquad (x)$$

where $E_{1T}$ and $E_{2H}$ are the total energies of the 2D monolayer Pt-MoSe$_2$ in 1T and 2H phases, respectively. In the present study, the value of $\Delta E_{1T-2H}$ is about +0.44 eV, which clearly indicates that the 2H phase is more stable than 1T phase of the 2D monolayer Pt-MoSe$_2$ TMD. Therefore, it can say that there is no phase transformation occurred (from the 2H phase to 1T phase of the 2D monolayer Pt-MoSe$_2$) after the single Pt atom doping in the pristine 2D MoSe$_2$ TMD (2x2 unite cell).

The electronic band structure and DOS calculations have been performed on the equilibrium structure of the 2D monolayer Pt-MoSe$_2$ material and the present periodic DFT-D calculations show that it has no band gap as depicted in Fig. 4b and a large electronic density is found around the Fermi energy level ($E_F$) as shown in Fig. 4c. This electron density of states around Fermi energy level was appeared due to the contribution of the electrons from *5d*-subshells of the Platinum (Pt) atom. This was confirmed by calculating the *d*-subshell contribution of the DOS of the Platinum which is given in Fig. 4d. We have plotted the electronic band structure of the 2D monolayer Pt-MoSe$_2$ material in high symmetric ***Γ-M-K-Γ*** direction as shown in Fig. 4b. Fermi energy level is located at -5.5 eV, and the number of four valence bands and four conduction bands were shown around the Fermi level ($E_F$). From the band structure calculations, we can see that one band (highlighted in blue color) is crossing the Fermi level from the conduction band to the valence band. Therefore, we can say that the bands are overlapped around the Fermi energy level and the 2D monolayer Pt-MoSe$_2$ material has metallic character with zero band gap. The total density of states (DOS) calculations of the 2D monolayer Pt-MoSe$_2$ depicts that the electronic density presents around the Fermi energy level ($E_F$) because of the significant hybridization of the Mo *3d*, Se *4p*, and Pt *5d* orbitals which was not observed in the case of the pristine 2D monolayer MoSe$_2$. The contributing component of the *5d*-orbitals electron density of states of the Pt atom in the total DOS has been computed as shown in Figure 4d and the present computation reveals that a large electron density appears at the $E_F$ due to the *5d*-orbitals electron density of states of the Pt atom after doping it in the 2D monolayer MoSe$_2$ material. We further extended our study by changing the Pt doping concentration in a 3x3 super cell of the 2D monolayer MoSe$_2$ which results in opening of the band gap by a small amount of 0.31 eV as the Pt-concentration has been decreased in this case. Thus, it is confirmed that the metallic behavior in the 2x2 supercell of Pt-MoSe$_2$ was appeared due to the doping of the Pt atom which makes the material conducting and hence it can be used as an efficient electrode for the better performance of both ORR and HER. However, the electronic properties calculations of the Pt-doped MoSe$_2$ material could not able to proof the performance of both ORR and HER.

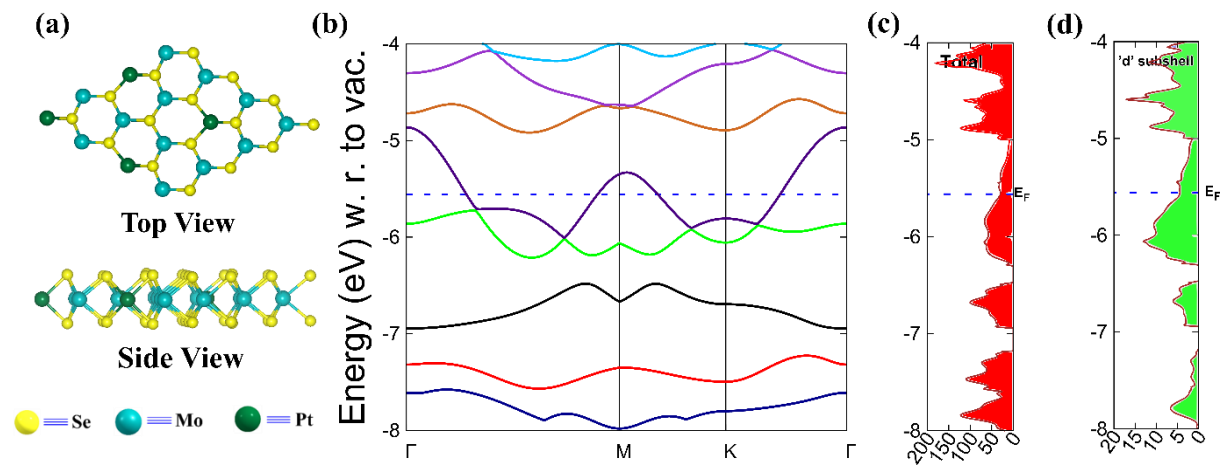

**Fig. 4.** (a) Top view and side view; (b) band structure; (c) total density of states of 2D monolayer Pt-MoSe$_2$ material and (d) *d*-subshell DOS of Pt atom towards the 2D monolayer Pt-MoSe$_2$ material is shown here.[7]

**Table 1.** Equilibrium structural parameters of pristine 2D monolayer MoSe$_2$ and 2D monolayer Pt-MoSe$_2$ (2x2 super cell) materials are reported here.

| System | Lattice parameters in Å | Interfacial angles in degree (º) | Space group and Symmetry | Electronic band gap ($E_g$) in eV | Average bond distance between atoms in Å | |
|---|---|---|---|---|---|---|
| | | | | | Mo-Se | Pt-Se |
| MoSe$_2$ | $a = b = 3.234$ | $\alpha = \beta = 90°$ and $\gamma = 120°$ | ***P-6m2*** | 2.21 | 2.506 | ------- |
| Pt doped MoSe$_2$ | $a = b = 6.738$ | $\alpha = \beta = 90°$ and | ***P1*** | 0.00 | 2.518 | 2.623 |

| (2x2 supercell) | | γ = 121.7° | | | | |
|---|---|---|---|---|---|---|
| | | | | | | |

To investigate the HER performance of the pristine 2D monolayer MoSe$_2$ material, we have developed a finite non-periodic cluster model system of the Mo$_{10}$Se$_{21}$. Similar way to study both the HER and ORR performances of the 2D Pt-MoSe$_2$ material, we developed a finite molecular cluster model system Pt$_1$-Mo$_9$Se$_{21}$ (i.e., Platinum doped 2D MoSe$_2$) and explored the reaction pathways, mechanism, thermodynamics, chemical kinetics, transition state (TS) structures, intermediates, and reaction barriers. M06-L DFT method was used to obtain the equilibrium geometries, reaction kinetics, thermodynamics, reaction intermediates and transition state structures (reaction barriers). In other words, to enable the use of the most accurate DFT for reaction barriers while describing solvation effects at the PCM level, we describe the Mo-edge of MoSe$_2$ using a Mo$_{10}$Se$_{21}$ cluster model and similarly Pt-/Mo-edge of the Pt-MoSe$_2$ using a Pt$_1$-Mo$_9$Se$_{21}$ finite non-periodic molecular cluster. This allows us to consider the introduction of protons and electrons separately and report free energies as a function of electrochemical potential and pH.

**Hydrogen evolution reaction (HER).** We have calculated the hydrogen adsorption free energy $\Delta G_H$ at the Se site of both the non-periodic molecular cluster model system Mo$_{10}$Se$_{21}$ and Pt$_1$-Mo$_9$Se$_{21}$ TMDs as it is an important descriptor to define the HER activity of the catalyst. The reactant supply and the product delivery must be fast for an electrocatalyst to be effective for HER. Therefore, the adsorption and desorption of the H atom should be strong on the surface of the electrocatalyst but getting both the qualities (strong adsorption and strong desorption) is very difficult. So, achieving a moderate behavior of H adsorption/desorption on the surface of electrocatalyst can enhance HER activity. Negative/positive values of the $\Delta G_H$ imply over-/under-binding of the H atoms to the catalyst surface, leading to non-optimal HER activity. The calculated hydrogen binding free energies on the pristine 2D monolayer MoSe$_2$ was found to be 0.41 eV. After the doping the Pt atom, the H adsorption becomes thermodynamically more favorable ($\Delta G_H$= -0.84 eV).

At first, we have carried out our investigations on the reaction mechanism of HER on the surfaces of pristine 2D monolayer MoSe$_2$ material. A detailed reaction mechanism with the proposed Volmer-Heyrovsky reaction pathway as the most promising pathway for the H$_2$ evolution is given below in Fig. 5. We next describe the predicted energetics for the various reaction steps relevant for HER. Using the cluster model, we can now add or subtract electrons and protons independently in discrete steps. First, we calculate the free energies of the most likely intermediates to serve as a basis for describing the thermodynamics of HER. Then, we examine the barriers of the various reaction steps to locate the rate limiting step of the reaction for $H_2$ evolution. Reaction free energies were calculated at each intermediate reaction steps listed in Table 2. The change in free energies values (ΔG) of the reaction steps are negative (except transition states (TSs)). This negative value of change of free energy tells the feasibility of the reaction steps i.e., all the reaction intermediates are stable and thermodynamically favorable. By employing the finite molecular cluster model $Mo_{10}Se_{21}$, we have incorporated no. of electrons and protons independently in discrete steps i.e., each step of the $H_2$ evolution reaction.

The absorption of an electron in the 2D monolayer $MoSe_2$ cluster at the SHE conditions took place which initiates the hydrogen evolution reaction (HER), leading to a negatively charged cluster $[MoSe_2]^-$. In the next step, hydride ion is absorbed on an energetically favorable Se site. This accompanies a hydride ($H^-$) shift to the reactive site of the catalyst. Here, it is the next adjoining metal atom (more specifically Mo) site. The detailed discussion of the HER mechanism on the surface of $Mo_{10}Se_{21}$ cluster has been given in the various steps below.

**Step I.** First reduction takes place on the surface of the 2D pristine $[MoSe_2]$ due to absorption of one electron ($e^-$) resulting $[MoSe_2]^-$ solvated in water. The addition of electron has been given by the curved arrow in Fig. 5. The first reduction potential to reduce $[MoSe_2]$ to $[MoSe_2]^-$ is about -898.9 mV. Here, the free energy of electron ($G(e^-)$) is calculated from the free energy of the half of hydrogen molecule ($G(1/2\ H_2)$) and the free energy of the proton ($G(H^+)$) as, $G(e^-) = G\left(\frac{1}{2}H_2\right) - G(H^+)$.

**Step II.** Adding one proton ($H^+$) to the active Se-edge of the 2D $[MoSe_2]^-$ results $[MoSe_2]H_{Se}$ (where the subscript Se at H represents that the hydrogen is adsorbed at the edge of the Se atom in the 2D $MoSe_2$). This step has been represented by the curved arrow with the addition of proton in Fig. 5. The first protonation of $[MoSe_2]^-$ to obtain $[MoSe_2]H_{Se}$ intermediate is done with an energy cost of ΔG = 6.13 kcal.mol$^{-1}$ as reported in Table 2.

**Step III.** Second reduction takes place at the 2D [MoSe$_2$]H$_{Se}$ with the addition of further one more electron to it which results [MoSe$_2$]H$_{Se}^-$. The second reduction is achieved with the second reduction potential about -572 mV.

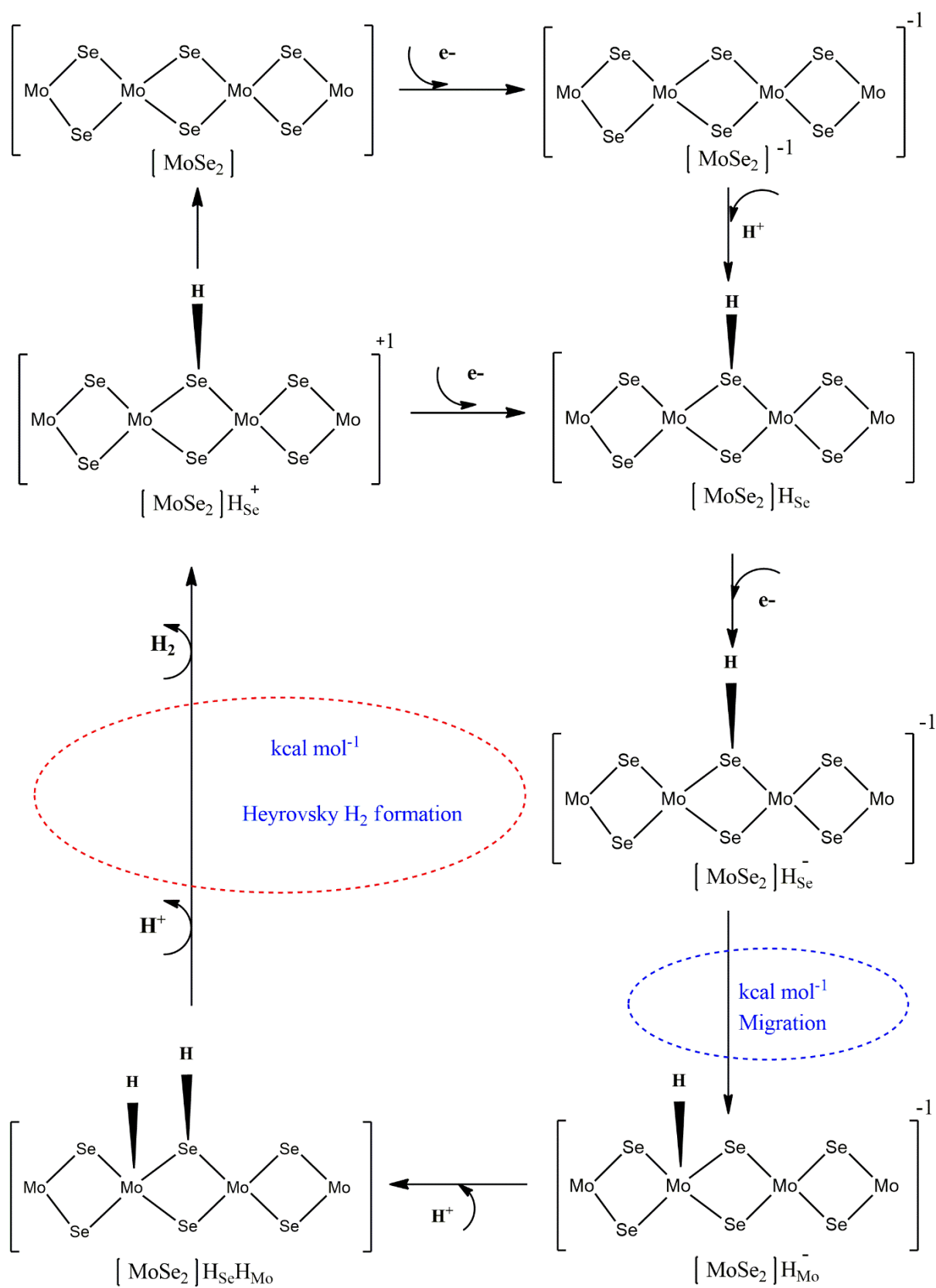

**Fig. 5.** HER mechanism with Volmer-Heyrovsky reaction pathway occurring on the active surfaces of pristine 2D monolayer MoSe$_2$ as an electrocatalyst; migration of H· indicates first Volmer transition

state (TS1) highlighted in blue color dotted ellipse with a barrier about 14.45 kcal.mol$^{-1}$ and the formation of H$_2$ in the Heyrovsky step gives the second transition state (TS2) highlighted by red color dotted ellipse with a barrier 21.77 kcal.mol$^{-1}$.

**Step IV.** In this step, the hydride atom (H$^{\bullet}$) adsorbed at the Se-site with a negative charge on it ([MoSe$_2$]H$_{Se}^{-}$) migrates to the nearest transition metal atom site (i.e., Mo-site) giving first transition state (TS1) commonly known as H$^{\bullet}$-migration TS or Volmer TS. This Volmer TS (TS1) is identified to have a single imaginary frequency of vibration with the mode of vibration. The activation barrier for this Volmer TS1 of the 2D monolayer pristine MoSe$_2$ is about $\Delta G$ = 13.44 – 14.45 kcal.mol$^{-1}$ in the gas phase and solvent phase calculations reported in Table 2.

**Step V.** [MoSe$_2$]H$_{Mo}^{-}$ results from the Volmer transition state (TS), noted by TS1, with an energy cost of $\Delta G$ = 4.96 kcal.mol$^{-1}$. Here, the subscript Mo at the H and the negative charge over it indicate that the proton is adsorbed at the Mo-atom site and the system is negatively charged with a single electron over it.

**Step VI.** Further, second protonation takes place at the Se-site of the [MoSe$_2$]H$_{Mo}^{-}$ resulting [MoSe$_2$]H$_{Mo}$H$_{Se}$ (subscripts H of the Mo and Se indicate that one hydrogen as adsorbed to the Mo atom and the another hydrogen is adsorbed to the Se-atom) with an energy cost of $\Delta G$ = 8.06 kcal.mol$^{-1}$.

**Step VII.** The second transition state called Heyrovsky TS, noted by TS2, results from [MoSe$_2$]H$_{Mo}$H$_{Se}$ in which a hydronium water cluster (3H$_2$O + H$_3$O$^{+}$) is added to the [MoSe$_2$]H$_{Mo}$H$_{Se}$ so that H$^{\bullet}$ at the Mo-site and H$^{+}$ from the hydronium water cluster recombine to evolve as H$_2$. The activation energy barrier for this Heyrovsky TS2 is found to be $\Delta G$ = 20.01 – 21.77 kcal.mol$^{-1}$ in the gas phase and solvent phase calculations as reported in Table 2.

**Step VIII.** After forming the Heyrovsky TS2, the system becomes to [MoSe$_2$]H$_{Se}^{+}$ with the successive evolution of one H$_2$ molecule along with the water cluster with an energy cost of $\Delta G$ = 41.40 kcal.mol$^{-1}$. Succeeding this HER process, the intermediate [MoSe$_2$]H$_{Se}^{+}$ either moves back to its initial phase of the pristine [MoSe$_2$] with the removal of the proton (H$^{+}$) absorbed at the Se-site or may be brought to the [MoSe$_2$]H$_{Se}$ with the addition of one further electron (e$^{-}$) as shown in the Volmer-Heyrovsky reaction mechanism pathway depicted in Fig. 5. The variations of energies ($\Delta E$, $\Delta H$ and $\Delta G$) with the proceeding of the HER on the surface of the 2D monolayer pristine [MoSe$_2$] are tabulated in the Table 2.

In summary, we calculated the two transition structures TS1 and TS2 (H$^•$-migration or Volmer TS and Heyrovsky TS in the case of the pristine 2D monolayer MoSe$_2$) during HER, the first one is at the time of H$^•$ migration from the selenium (Se) site to the metal site (here Mo). The second TS has been calculated at the step where the H$_2$ molecule is formed during the Heyrovsky reaction step in the HER process at the surface of pristine 2D monolayer MoSe$_2$ material. Both the Volmer and Heyrovsky reaction barriers are 13.44 and 20.01 kcal.mol$^{-1}$, respectively, computed in gas phase. The PCM calculations have been carried out by taking water as solvent implicitly with the dielectric constant of 78.35. These Volmer and Heyrovsky energy barriers in solvent phase were observed to be 14.45 and 21.77 kcal.mol$^{-1}$, respectively. Therefore, the rate-determining step of this HER process is Heyrovsky's reaction step as the Heyrovsky's reaction barrier energy (TS2) is higher than H$^•$-migration reaction barrier (TS1). The equilibrium geometries of the reactant, products, intermediates, and TSs occurred in HER process on the surfaces of MoSe$_2$ are shown in Fig. 6a-i.

**Table 2.** Changes of various energy values ($\Delta E$, $\Delta H$ and $\Delta G$) during the HER process on the surface of the pristine 2D monolayer MoSe$_2$ material is reported here. The units of the energies are expressed in kcal.mol$^{-1}$.

| Sl. No. HER steps | HER Reaction Steps | $\Delta E$ kcal.mol$^{-1}$ (Gas Phase) | $\Delta H$ kcal.mol$^{-1}$ (Gas Phase) | $\Delta G$ kcal.mol$^{-1}$ (Gas Phase) |
|---|---|---|---|---|
| Step I | [MoSe$_2$] → [MoSe$_2$]$^−$ | -19.26 | -19.34 | -20.73 |
| Step II | [MoSe$_2$]$^−$ → [MoSe$_2$]H$_{Se}$ | -10.39 | -5.19 | -6.13 |
| Step III | [MoSe$_2$]H$_{Se}$ → [MoSe$_2$]H$_{Se}^−$ | -14.46 | -14.48 | -13.19 |
| Step IV | [MoSe$_2$]H$_{Se}^−$ → TS1 | 14.69 | 15.08 | 13.44 |
| Step V | TS1 → [MoSe$_2$]H$_{Mo}^−$ | -4.87 | -3.24 | -4.96 |
| Step VI | [MoSe$_2$]H$_{Mo}^−$ → [MoSe$_2$]H$_{Mo}$H$_{Se}$ | -13.49 | -8.27 | -8.06 |

| Step VII | [MoSe$_2$]H$_{Mo}$H$_{Se}$ → TS2 | 35.27 | 34.84 | 20.01 |
| Step VIII | TS2 → [MoSe$_2$]H$_{Se}$$^+$ | -47.99 | -48.22 | -41.40 |

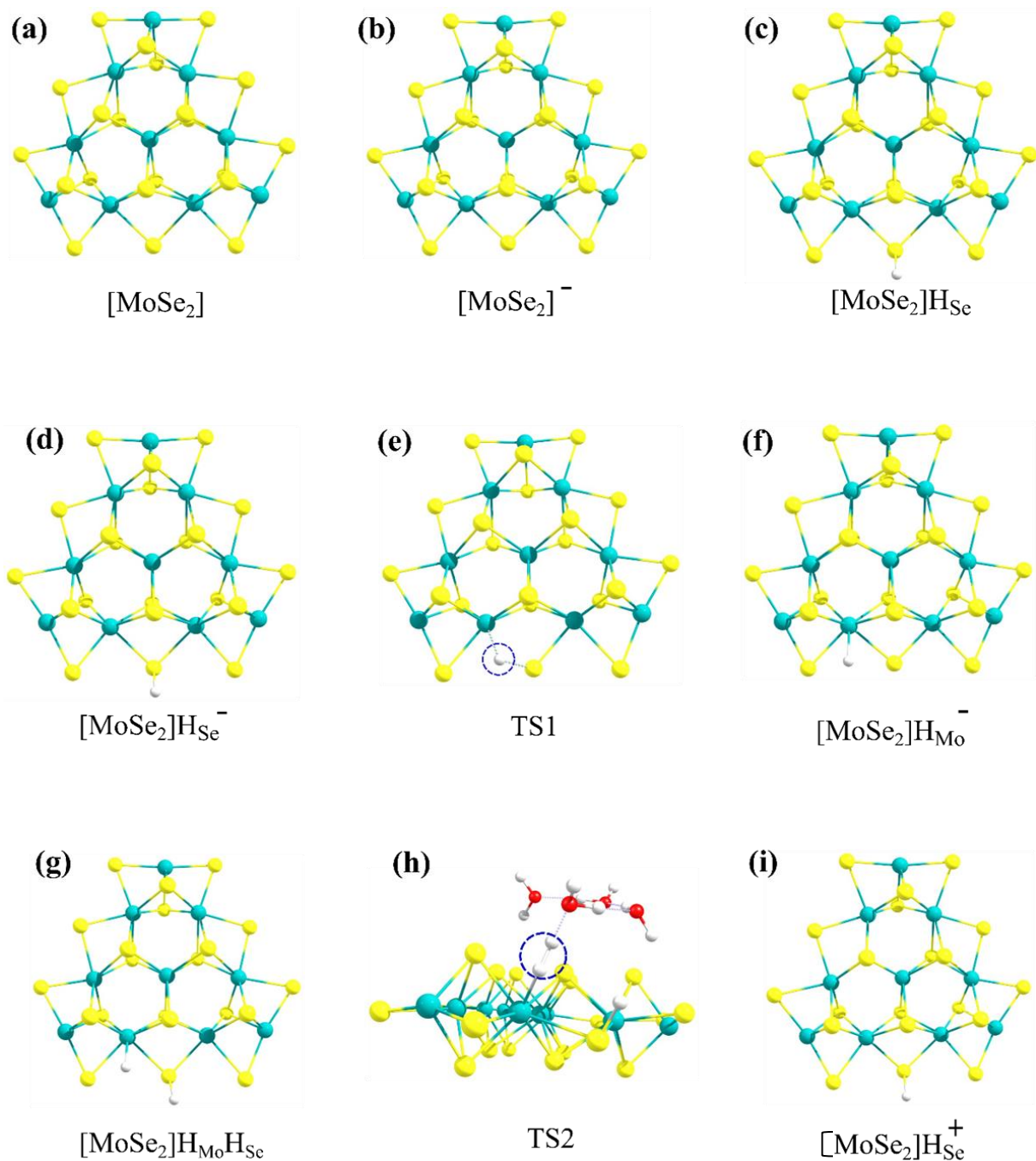

(a) [MoSe$_2$]  (b) [MoSe$_2$]$^-$  (c) [MoSe$_2$]H$_{Se}$

(d) [MoSe$_2$]H$_{Se}$$^-$  (e) TS1  (f) [MoSe$_2$]H$_{Mo}$$^-$

(g) [MoSe$_2$]H$_{Mo}$H$_{Se}$  (h) TS2  (i) [MoSe$_2$]H$_{Se}$$^+$

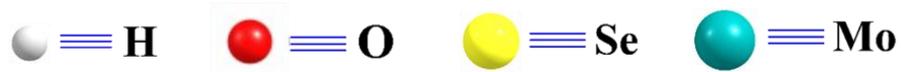

○ = H   ● = O   ● = Se   ● = Mo

**Fig. 6.** Equilibrium geometries of (a) **[MoSe$_2$]**; (b) **[MoSe$_2$]$^-$**; (c) **[MoSe$_2$]H$_{Se}$**; (d) **[MoSe$_2$]H$_{Se}^-$**; (e) **TS1**; (f) **[MoSe$_2$]H$_{Mo}^-$**; (g) **[MoSe$_2$]H$_{Se}$H$_{Mo}$**; (h) **TS2** and (i) **[MoSe$_2$]H$_{Mo}^+$** computed by the M06-L DFT method considering a molecular cluster model system **Mo$_{10}$Se$_{21}$** to represent 2D monolayer **MoSe$_2$** are shown here.

To study the HER mechanisms and electrocatalytic activity of the 2D monolayer Pt-MoSe$_2$ material, we have computationally explored the reaction pathways of the H$_2$ evolution process by considering Volmer-Heyrovsky mechanisms as the most promising pathway for HER. This Volmer-Heyrovsky mechanism has two crucial steps; (i) Volmer reaction mechanism step which involves the migration of H$^\bullet$ from selenium (chalcogen) site to platinum (metal) site and (ii) Heyrovsky step which involves the evolution of H$_2$ by the recombination of one solvated proton from the water (solvent used here) with the interaction of absorbed H at the Pt-atom of the 2D Pt-MoSe$_2$. A detailed schematic diagram of each step involved during the Volmer-Heyrovsky mechanism pathway of HER for Pt-MoSe$_2$ is given in the Fig. 7. A molecular cluster model of the Pt$_1$-Mo$_9$Se$_{21}$ system for the 2D Pt-MoSe$_2$ material has been proposed here to study the HER reaction mechanism and the reaction energy barriers (both the Volmer and Heyrovsky Transition States). The equilibrium geometries of the reactant, products, intermediates, and TSs appeared in HER process on the surfaces of Pt-MoSe$_2$ are shown in Fig. 8a-i.

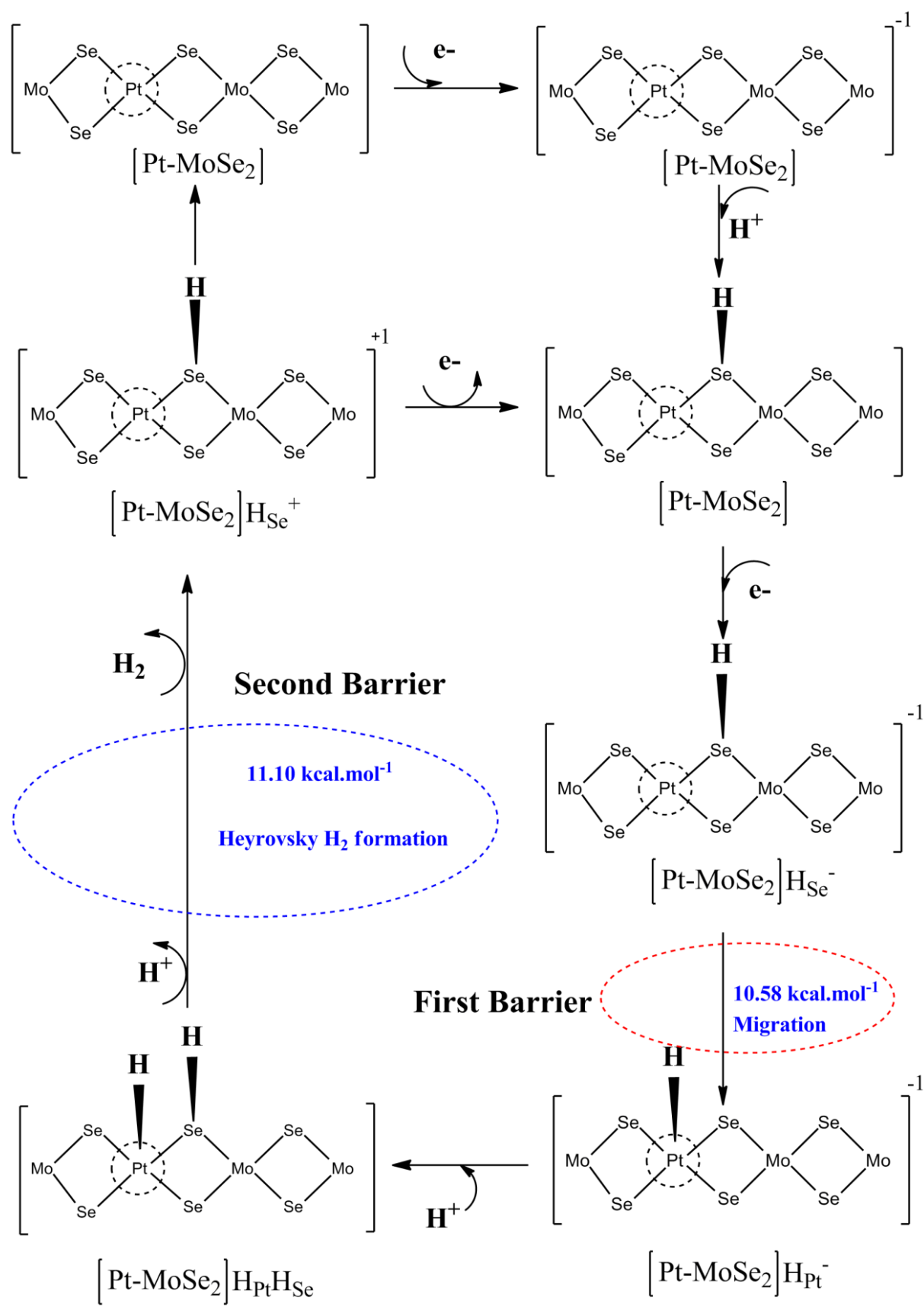

**Fig. 7.** HER mechanism with the Volmer-Heyrovsky pathway occurring on the surfaces of platinum doped 2D monolayer MoSe$_2$ as an electrocatalyst; migration of H$^\bullet$ indicates first Volmer TS (represented by TS3) highlighted in red color dotted ellipse with a barrier about 10.58 kcal.mol$^{-1}$ and the formation of H$_2$ in the Heyrovsky step gives the second TS (represented by TS4) highlighted by blue color dotted ellipse with a barrier 11.10 kcal.mol$^{-1}$.

The HER process takes place on the active sites (i.e. Pt-doping sites) in the 2D monolayer Pt-MoSe$_2$ through the Volmer-Heyrovsky reaction mechanism as shown in the proposed reaction pathway in Fig. 7. This whole mechanism is carried out with the following steps with the discussion of Gibbs free energy changes.

**Step I.** HER is started at the SHE conditions (E = 0 V and pH = 0) with the initial stage of our [Pt-MoSe$_2$] system of interest. [Pt-MoSe$_2$]$^-$ intermediate results from [Pt-MoSe$_2$] with the addition of a single electron (e$^-$) to the bare catalyst. The first reduction (i.e., the reduction of the [Pt-MoSe$_2$] to [Pt-MoSe$_2$]$^-$) has been carried out in this study and the first reduction potential is about -1.337 V, where the free energy of electron (G(e$^-$)) is calculated with the same procedure as discussed in the case of the 2D pristine [MoSe$_2$] material.

**Step II.** First proton (H$^+$) mostly prefers to be adsorbed at the most active Se-edge of the [Pt-MoSe$_2$]$^-$ intermediate near to the Pt doped atom. Due to the first protonation, the system becomes [Pt-MoSe$_2$]H$_{Se}$ (where the subscript Se at H represents that the hydrogen is adsorbed at the Se atom of the 2D monolayer of Pt-MoSe$_2$) with an energy cost of ΔG = 5.52 kcal.mol$^{-1}$.

**Step III.** With the addition of second electron to the [Pt-MoSe$_2$]H$_{Se}$ intermediate the catalyst reduces to [Pt-MoSe$_2$]H$_{Se}^-$. This second reduction potential is obtained by the same DFT method, and the second reduction potential is about -283.17 mV.

**Step IV.** In the next step of the HER process, the hydride atom (H$^\bullet$) adsorbed at the Se-site in the [Pt-MoSe$_2$]H$_{Se}^-$ intermediate migrates to the nearest Pt atom site yielding the first transition state (represented by TS3) commonly known as H$^\bullet$-migration TS or Volmer TS. A harmonic vibrational analysis has been performed to identify the TS. This Volmer TS (i.e., TS3) has a single imaginary frequency of vibration at the site where the H$^\bullet$ is migrating towards the Platinum from the Se site as shown in Fig. 8e. The activation barrier for this H$^\bullet$-migration TS3 of the 2D Pt-MoSe$_2$ is about ΔG = 9.29

kcal.mol$^{-1}$ in the gas phase and 10.55 kcal.mol$^{-1}$ in the solvent phase calculations which are lower than the same H$^{\bullet}$-migration TS1 of the 2D pristine MoSe$_2$. In other words, one important observation here is that the energy barriers corresponding to the Volmer TS3 of 2D [Pt-MoSe$_2$] is less than that of 2D [MoSe$_2$] which indicates that the [Pt-MoSe$_2$] material is a good electrocatalyst for H$_2$ evolution. This calculation shows that the barrier of the TS3 in the case of 2D Pt-MoSe$_2$ is about 4.15 kcal.mol$^{-1}$ lower than the TS1 in the case of the pristine 2D MoSe$_2$.

**Step V.** Through the Volmer TS3, the catalyst becomes [Pt-MoSe$_2$]H$_{Pt}^{-}$ with an energy cost about $\Delta G = -59.00$ kcal.mol$^{-1}$. Here, the subscript Pt at the H and the negative charge over it indicate that the proton is adsorbed at the Pt-atom and the system is negatively charged with a single electron over it, respectively. The equilibrium structure of this intermediate is shown in Fig. 8f.

**Step VI.** Further second protonation takes place at the Se-atom of the [Pt-MoSe$_2$]H$_{Pt}^{-}$ intermediate resulting [Pt-MoSe$_2$]H$_{Pt}$H$_{Se}$ with an energy cost of $\Delta G = -56.78$ kcal.mol$^{-1}$ indicating that the reaction is thermodynamically favorable. The changes of electronic energy, free energies and enthalpies of various reaction steps are reported in Table 3.

**Table 3.** Comparison of the changes of various energy values ($\Delta E$, $\Delta G$ and $\Delta H$ expressed in kcal.mol$^{-1}$) of various intermediates resulted during HER process on the surfaces of the 2D monolayer Pt-MoSe$_2$ catalyst in the gas phase calculation are reported here

| Sl. No. HER steps | HER Reaction Steps | $\Delta E$ kcal.mol$^{-1}$ (Gas Phase) | $\Delta H$ kcal.mol$^{-1}$ (Gas Phase) | $\Delta G$ kcal.mol$^{-1}$ (Gas Phase) |
|---|---|---|---|---|
| Step I | [Pt-MoSe$_2$] → [Pt-MoSe$_2$]$^{-}$ | -29.60 | -29.66 | -30.83 |
| Step II | [Pt-MoSe$_2$]$^{-}$ → [Pt-MoSe$_2$]H$_{Se}$ | -9.00 | -4.06 | -5.52 |
| Step III | [Pt-MoSe$_2$]H$_{Se}$ → [Pt-MoSe$_2$]H$_{Se}^{-}$ | -7.68 | -7.82 | -6.53 |
| Step IV | [Pt-MoSe$_2$]H$_{Se}^{-}$ → TS3 | 8.12 | 7.57 | 9.29 |
| Step V | TS3 → [Pt-MoSe$_2$]H$_{Pt}^{-}$ | -58.74 | -57.98 | -59.00 |
| Step VI | [Pt-MoSe$_2$]H$_{Pt}^{-}$ → [Pt-MoSe$_2$]H$_{Pt}$H$_{Se}$ | -60.27 | -55.23 | -56.78 |

| Step VII | [Pt-MoSe$_2$]H$_{Pt}$H$_{Se}$ → TS4 | 6.56 | 5.31 | 10.58 |
| Step VIII | TS4 → [Pt-MoSe$_2$] H$_{Se}^+$ | -27.37 | -26.99 | -39.92 |

**Step VII.** In this step, the [Pt-MoSe$_2$]H$_{Pt}$H$_{Se}$ intermediate interact with the water cluster (3H$_2$O_H$_3$O$^+$) to form the H$_2$ followed by the Volmer−Heyrovsky mechanism for HER process. The Volmer-Heyrovsky mechanism is more complicated because it is necessary to solvate the H$_3$O$^+$ source of the proton (H$^+$) along the reaction pathway. In this reaction mechanism a TS has been formed during the reaction known as Heyrovsky transition state noted by TS4. This Heyrovsky transition state (TS4) results from the [Pt-MoSe$_2$]H$_{Pt}$H$_{Se}$ in which a hydronium water cluster (3H$_2$O_H$_3$O$^+$) is brought into the vicinity of [Pt-MoSe$_2$]H$_{Pt}$H$_{Se}$ so that H$^•$ at the Pt-site and H$^+$ from the hydronium water cluster recombine to evolve as H$_2$. The equilibrium geometry of the TS4 is shown in Fig. 8h, and it is observed that this Heyrovsky TS4 has a negative frequency while optimizing the TS. From our present DFT calculations, it was found that the activation energy barrier for this Heyrovsky TS2 is ΔG = 10.58 kcal.mol$^{-1}$ in the gas phase and 11.10 kcal.mol$^{-1}$ in the solvent phase calculations, respectively. We have observed that to obtain accurate results it requires the use of a cluster of 4 waters, one of which is protonated at the beginning but all of which are neutral at the end. However, in the reaction between an adsorbed hydrogen atom and the hydronium bound proton, the water cluster must rearrange to expose the proton for reaction. The Heyrovsky reaction barrier in this process is much lower than the Se-H case, making them the most favorable transition structures to form H$_2$. The Heyrovsky barrier in the case of [Pt-MoSe$_2$] is about 11.19 kcal.mol$^{-1}$ lower than the case of the pristine 2D [MoSe$_2$] which indicates that the 2D [Pt-MoSe$_2$] is an excellent electrocatalyst for H$_2$ evolution reaction.

**Step VIII.** The catalyst becomes an intermediate [Pt-MoSe$_2$]H$_{Se}^+$ after the formation of Heyrovsky TS4 with the successive evolution of one H$_2$ molecule and the water cluster with an energy cost of ΔG = 39.92 kcal.mol$^{-1}$. Then, after forming the intermediate [Pt-MoSe$_2$]H$_{Se}^+$, it may become [Pt-MoSe$_2$] with the removal of the proton (H$^+$) absorbed at the Se-site or may form [Pt-MoSe$_2$]H$_{Se}$ with the addition of one further electron (e$^-$) as shown in the Volmer-Heyrovsky reaction mechanism pathway in Fig. 7. All the energy changes during this reaction step computed in the gas phase reactions of the HER process on the surface of the 2D monolayer [Pt-MoSe$_2$] material is tabulated in the Table 3, and all the equilibrium structures are depicted in Fig. 8.

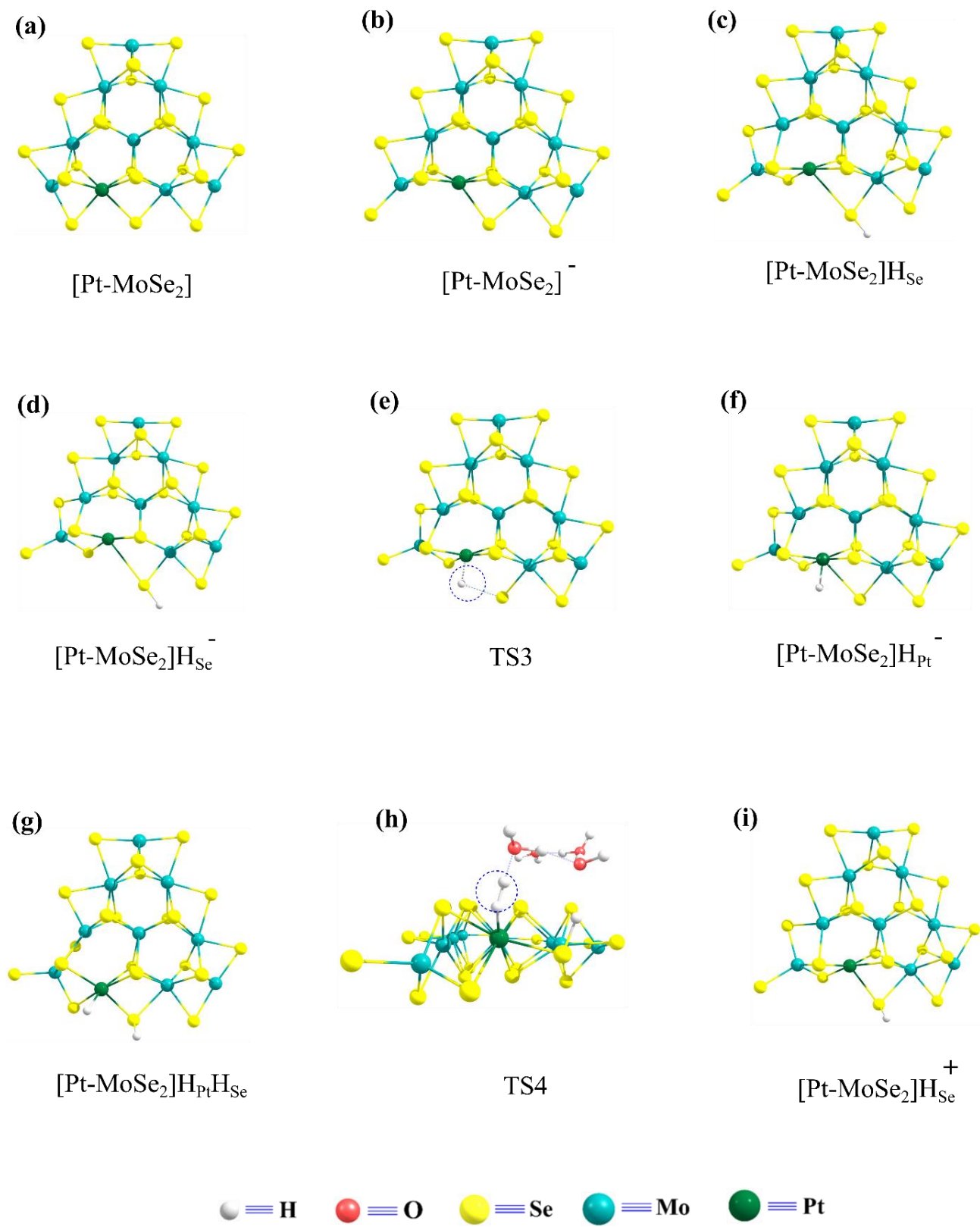

(a) [Pt-MoSe$_2$]  (b) [Pt-MoSe$_2$]$^-$  (c) [Pt-MoSe$_2$]H$_{Se}$

(d) [Pt-MoSe$_2$]H$_{Se}^-$  (e) TS3  (f) [Pt-MoSe$_2$]H$_{Pt}^-$

(g) [Pt-MoSe$_2$]H$_{Pt}$H$_{Se}$  (h) TS4  (i) [Pt-MoSe$_2$]H$_{Se}^+$

○ = H   ● = O   ● = Se   ● = Mo   ● = Pt

**Fig. 8.** Equilibrium geometries of (a) **[Pt-MoSe$_2$]**; (b) **[Pt-MoSe$_2$]$^-$**; (c) **[Pt-MoSe$_2$]H$_{Se}$**; (d) **[Pt-MoSe$_2$]H$_{Se}^-$**; (e) **TS3**; (f) **[Pt-MoSe$_2$]H$_{Pt}^-$**; (g) **[Pt-MoSe$_2$]H$_{Pt}$H$_{Se}$**; (h) **TS4** and (i) **[Pt-MoSe$_2$]H$_{Pt}^+$** are shown here computed by the M06-L DFT method considering a molecular cluster model system Pt$_1$-Mo$_9$Se$_{21}$ to represent 2D monolayer **Pt-MoSe$_2$** TMD.

In the present study, we have also explored the reaction mechanism of HER in the solvent phase studying the PCM model which is a commonly used method in computational material chemistry to model solvation effects as stated earlier. It represents one of the most successful examples among continuum solvation models and the solute (a single molecule or a cluster containing the solute and some relevant solvent molecules) is described quantum mechanically, while the solvent is approximated as a structureless continuum whose interaction with the solute is mediated by its relative permittivity, ε, i.e., dielectric constant. Here, the solvent phase calculations have been carried out by taking water as a solvent with the dielectric constant (ε) value of 80.13, which helps in the solvation effects in the HER mechanism of 2D monolayer Pt-MoSe$_2$. In Table 4, we have listed the reaction barriers (Volmer, i.e. H$^\bullet$-migration and Heyrovsky) of different systems studied in both the gas and solvent phases. In the case of the 2D monolayer pristine MoSe$_2$ TMD, H$^\bullet$-migration reaction barriers are 13.44 kcal.mol$^{-1}$ and 14.45 kcal.mol$^{-1}$ computed in the gas and solvent phases, respectively. Similarly, the H$^\bullet$-migration reaction barriers are found to be 9.29 kcal.mol$^{-1}$ and 10.55 kcal.mol$^{-1}$ obtained in gas and solvent phases, respectively, when the reaction occurs on the surfaces of the 2D monolayer Pt-MoSe$_2$ system. The Heyrovsky reaction barriers in both the gas phase and solvent phases are 20.01 kcal.mol$^{-1}$ and 21.77 kcal.mol$^{-1}$, respectively, when the HER occurs on the surfaces of the 2D pristine MoSe$_2$ material. Similarly, the Heyrovsky reaction barriers are in the range of 10.58 kcal.mol$^{-1}$ and 11.10 kcal.mol$^{-1}$, respectively, when the reaction occurs at the surfaces of the 2D monolayer Pt-MoSe$_2$ system. The activation energy barrier of the TS3 (in the H$^\bullet$-migration step during Volmer reaction step) in the solvent phase during HER on the surfaces of the Pt-MoSe$_2$ is about 10.55 kcal.mol$^{-1}$ which is about 7.75 - 9.95 kcal.mol$^{-1}$ amount lower than the pristine 2D MoS$_2$ and WS$_2$ materials reported in Table 4. This TS3 barrier in the case of the 2D monolayer Pt-MoSe$_2$ material is also lower than the pristine 2D MoSe$_2$ about 3.90 kcal.mol$^{-1}$ computed by the same level of theory. The second energy barrier due to the Heyrovsky reaction to form the TS4 in the solvent phase, i.e., Heyrovsky TS, is about 11.10 kcal.mol$^{-1}$ which is also lower than the energy barrier of the Heyrovsky step of the other TMD materials listed in Table 4.

Therefore, this Heyrovsky TS4 represents the rate-determining step of the HER. Both the reaction barriers i.e., TS3 and TS4 in the case of the 2D monolayer Pt-MoSe$_2$ material are reasonable accord with the W$_x$Mo$_{1-x}$S$_2$ alloys as reported in Table 4. Therefore, the present DFT study indicates that the 2D monolayer Pt-MoSe$_2$ material is an excellent electrocatalyst for effective H$_2$ evolution. A comparative listing of the values of turn over frequency (TOF) and these two energy barriers of various TMDs have been given in Table 4. This comparative result studied here clarifies and supports our approach towards the doping of platinum atom in the pristine 2D MoSe$_2$ which enhanced its electrocatalytic activity for HER mechanism with comparatively lower values of energy barriers.

**Table 4.** A Summary of the previously reported TMDs along with the present studied systems (2D monolayer MoSe$_2$ and 2D monolayer Pt-MoSe$_2$ materials) for the electrocatalytic HER with the values of the activation barriers and Turn Over Frequency (TOF).

| 2D TMDs | H·-migration reaction barrier kcal.mol$^{-1}$ (Gas Phase) / (Solvent Phase) | Heyrovsky reaction barrier kcal.mol$^{-1}$ (Gas Phase) / (Solvent Phase) | Turn Over Frequency (TOF) (sec$^{-1}$) | References |
|---|---|---|---|---|
| MoS$_2$ | 17.7 / 20.5 | 23.6 / 23.8 | 2.1x10$^{-5}$ | 8, 23 |
| WS$_2$ | 12.4 / 18.1 | 14.5 / 21.3 | 1.5x10$^{-3}$ | 9 |
| W$_x$Mo$_{1-x}$S$_2$ | 6.8 / 11.90 | 11.5 / 13.3 | 1.1x10$^3$ | 9 |
| MoSe$_2$ | 13.44 / 14.45 | 20.01 / 21.77 | 6.01x10$^{-4}$ | This work |
| Pt-MoSe$_2$ | 9.29 / 10.55 | 10.58 / 11.10 | 4.26x10$^4$ | This work. |

**HOMO-LUMO calculations for HER on both 2D MoSe$_2$ and 2D Pt-MoSe$_2$.** To explain the HER catalytic performance of both the pristine 2D monolayer MoSe$_2$ and 2D monolayer Pt-MoSe$_2$ materials, we performed NBO (Natural Bond Orbital), highest occupied molecular orbital (HOMO) and lowest unoccupied molecular orbital (LUMO) calculations for both the Volmer and Heyrovsky transition states TS1, TS2, TS3 and TS4. These calculations for the Heyrovsky transition state structures TS2 and TS4 can better explain the Heyrovsky mechanism (as the Heyrovsky's reaction step is the rate-determining step) for the effective HER from the perspective electronic charge cloud and overlapping of molecular orbitals during the formation of $H_2$. The optimized structures of both the TS1 and TS2 of the pristine 2D monolayer $MoSe_2$ and the TS3 and TS4 of the 2D monolayer Pt-$MoSe_2$ materials were taken into consideration for the study of HOMO and LUMO which is given in Fig. 9 and 10.

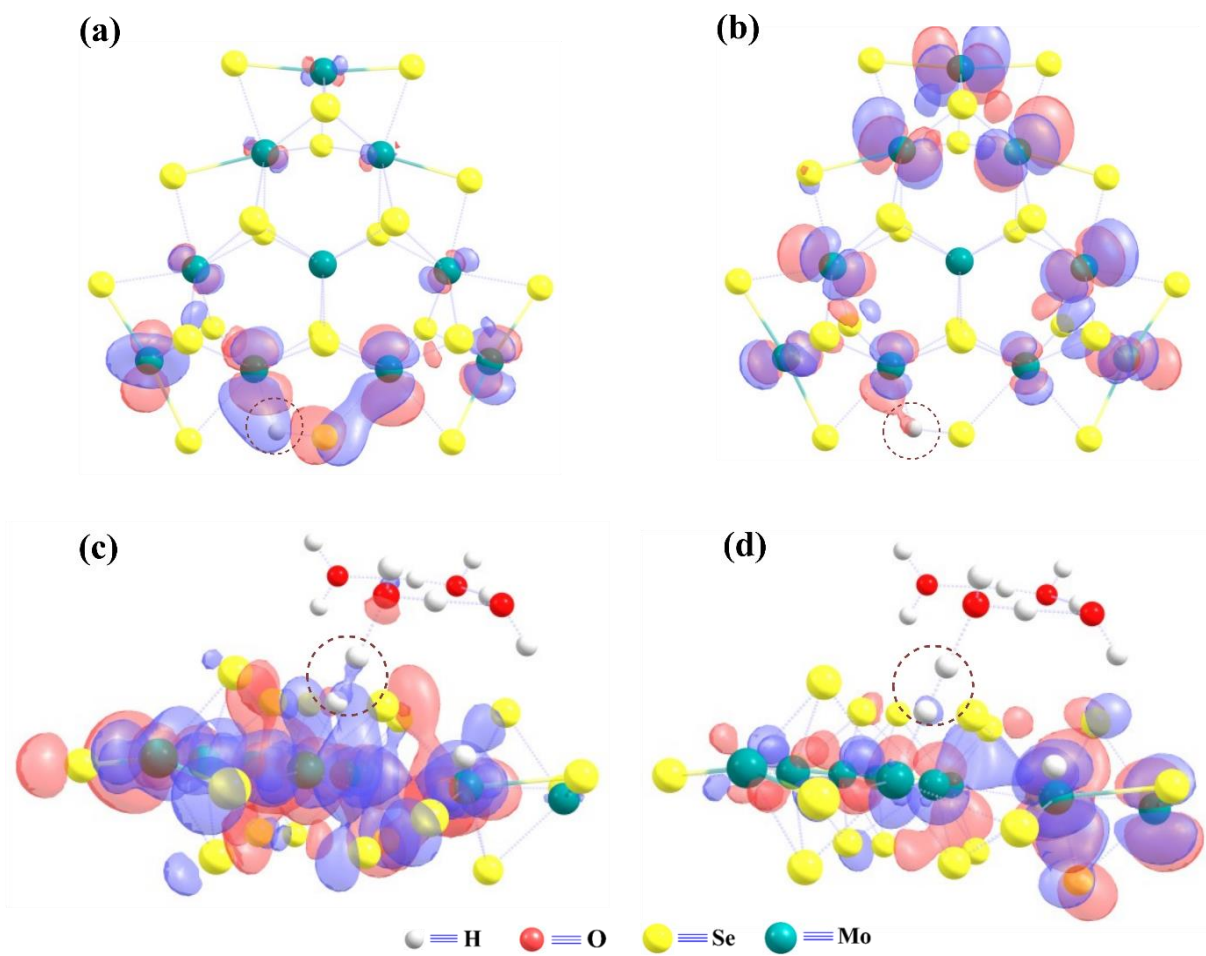

**Fig. 9.** (a) HOMO, (b) LUMO structure of the transition state TS1 occurred in the Volmer reaction step; and (c) HOMO, (d) LUMO structure of the transition state TS2 occurred in the Heyrovsky reaction step for pristine 2D monolayer $MoSe_2$ are shown here.

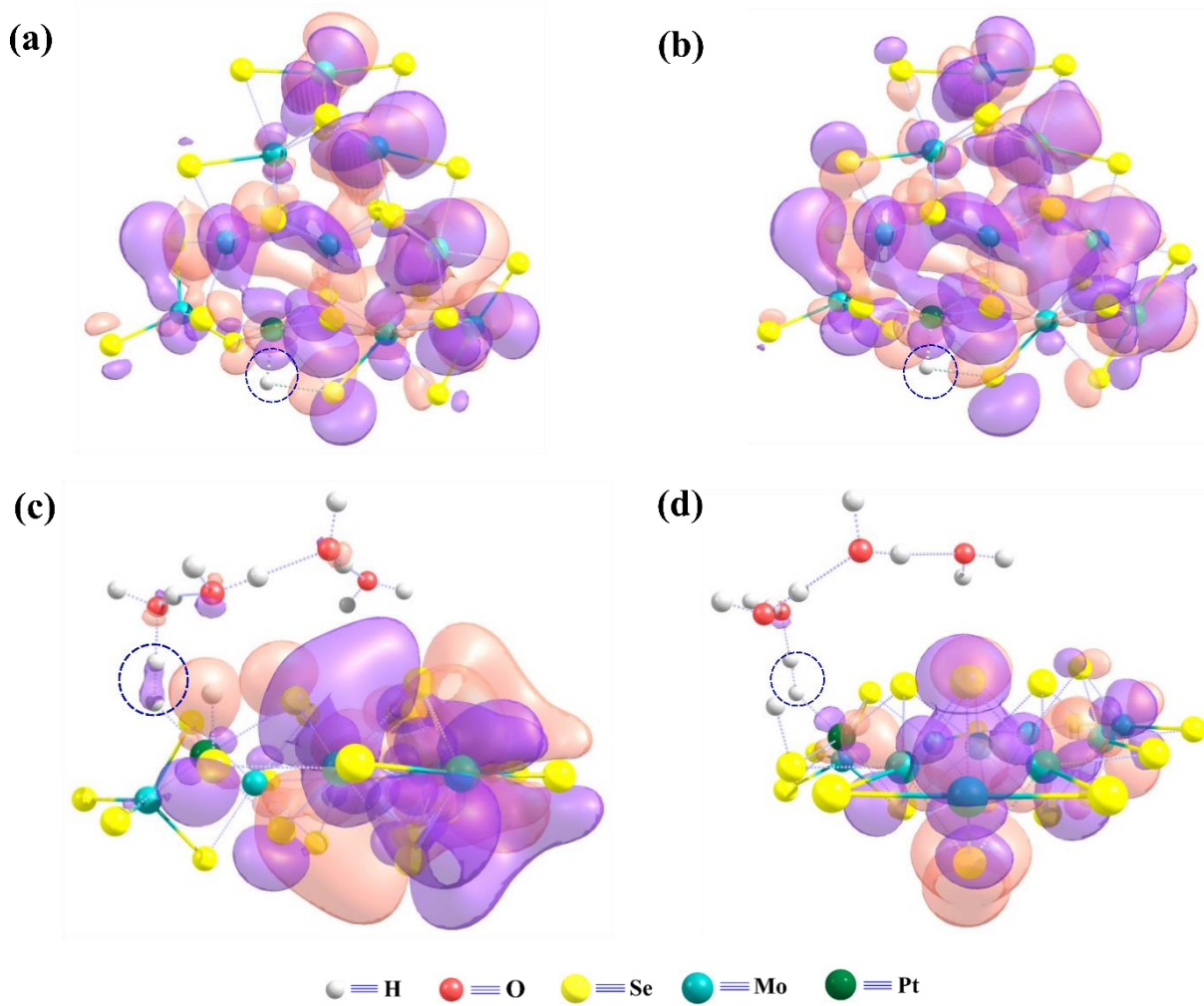

**Fig. 10.** (a) HOMO, (b) LUMO structures of the transition state TS3 occurred in the Volmer reaction step and (c) HOMO, (d) LUMO structures of the transition state TS4 occurred in the Heyrovsky reaction step for the 2D monolayer Pt-MoSe$_2$ are shown here.

  The NBO study provides a complete mathematical information about the binding orbitals for maximum possible energy of electron density.[65] The benefit of performing this study is that it provides the information of inter and intra molecular interactions. Second order Fock-matrix has been involved for the evaluation of donor acceptor interaction.[66] On the basis of HOMO and LUMO energies with the wave functions, reactivity and molecular stability can be understood easily. The energy difference of HOMO and LUMO study gives the stability of molecules. The capabilities of electron donor and acceptor unit is illustrated by HOMO energy and LUMO energy, respectively. This study found that the TS3 and TS4 in the case of the 2D Pt-MoSe$_2$ are alleviated by the overlapping of the *s*-orbital of the H$_2$, and the *d*-orbital of the Pt metal atom as displayed in Fig. 10a-d. Therefore, it can say that this better stabilization (compared to the pristine $MoS_2$, $WS_2$ and $MoSe_2$) of the orbitals in the reaction limiting step TS4 for $H_2$-formation is a key for reducing the Heyrovsky's reaction barrier, thus the overall catalysis indicating a better electrocatalytic performance for $H_2$ evolution. The electronic charged cloud represents the positive and negative part of the wave functions with the violet and pink color transparent cloud around Mo, Pt, Se, O, and H atoms as shown in Fig. 10. The electron cloud (noted by violet color) in between two H atoms in the Heyrovsky transition state structure highlighted by the blue dotted circle is more in the case of 2D Pt-$MoSe_2$ TMD in comparison to the pristine 2D $MoSe_2$ TMD structure in presence of four explicit water molecules in Heyrovsky step of $H_2$ evolution. Therefore, the Heyrovsky reaction step has been sustained by overlapping the atomic orbitals of two H atoms with the adjacent hydronium ($H_3O^+$) in the explicit water cluster to form the $H_2$ molecules during HER. It should be noted here that these overlapping with better stabilization (compared to the pristine 2D monolayer $MoS_2$ and $WS_2$) determines the reaction barriers and electrocatalytic performance of the 2D Pt-$MoSe_2$ TMD for effective HER. This is the reason behind the 2D Pt-$MoSe_2$ material which has shown an excellent catalytic activity for effective HER.

**Oxygen reduction reaction (ORR).** Pristine 2D monolayer $MoSe_2$ has a large electronic band gap ($E_g$) about 2.21 eV and lack of electron density at the Fermi energy ($E_F$) level due to which is does not have sufficient electronic conductivity. This obstructs electron transfer during the ORR. Therefore, the pristine 2D monolayer $MoSe_2$ cannot be used as an electrocatalyst for ORR. Doping the platinum atom reduces the band gap to 0 eV having a high electron density at the Fermi level ($E_F$) which in terms enhances the electronic conductivity of the overall Pt-$MoSe_2$ material as shown in Fig. 4c. Due to zero energy bandgap of the 2D monolayer Pt-$MoSe_2$ it can be adopted as an effective electrocatalyst for ORR like HER. The present study has focused on the 2D monolayer Pt-$MoSe_2$ material considering a $Pt_1$-$Mo_9Se_{21}$ finite non-periodic molecular cluster model system to investigate the ORR electrocatalytic activities. The detailed ORR mechanism on the active surface of the 2D Pt-$MoSe_2$ is schematically represented in Fig. 11.

The ORR has been started at the standard hydrogen electrode (SHE) conditions by adsorption of one $O_2$ molecule on the energetically favorable selenium (Se) site which in turn is attached to the Pt atom with a Gibbs free energy change ($\Delta G$) about 2.74 kcal.mol$^{-1}$ resulting [$O_2$_Pt-$MoSe_2$]. The equilibrium bond distances of the Pt-Se and Se-O were found to be 2.54 Å and the 2.30 Å, respectively. After $O_2$ adsorption, the bond length of $O_2$ molecule was stretched by an amount of 0.017 Å and one oxygen atom moved on the other Se site attached to the Pt atom with an energy cost of -8.11 kcal.mol$^{-1}$ giving the [2O_Pt-MoSe$_2$] intermediate as shown in Fig. 11 and the equilibrium geometries of the intermediates are shown in Fig. 12. The 2O formation step was achieved with negative value of free energy change so this step is thermodynamically favorable. The next step of the ORR was followed by the addition of one electron (e$^-$) coming from the anode side through the external circuit in the fuel cell forming [2O_Pt-MoSe$_2$]$^-$ with a free energy change of ΔG= -9.79 kcal.mol$^{-1}$. After the addition of one electron, one proton (H$^+$) coming through proton exchange membrane is attached to one of the oxygens with an energy cost of -5.36 kcal.mol$^{-1}$ giving rise to the [OH_O_Pt-MoSe$_2$] intermediate. The process of addition of electron and proton was repeated so that in the next step, the first transition state (represented by TS5) is resulted with an energy cost of 15.79 kcal.mol$^{-1}$. After the formation of TS5 transition state, one water molecule (H$_2$O) is removed so that one oxygen atom remains attached to the nearest Se site which in turns connected to the Pt atom resulting [O_Pt-MoSe$_2$] with the Gibbs free energy change about ΔG = -63.39 kcal.mol$^{-1}$.

The process of addition of electron and proton was repeated resulting [O_Pt-MoSe$_2$]$^-$ and [HO_Pt-MoSe$_2$] intermediates with the free energy changes of ΔG= -28.46 and -26.15 kcal.mol$^{-1}$, respectively. The addition process of one electron and furthermore one proton is performed so in the next step a second transition state represented by TS6 is formed with an energy cost of 7.71 kcal.mol$^{-1}$. In the last step one more H$_2$O molecule was removed so that finally the system comes to its initial Pt-MoSe$_2$ stage with an energy cost of -58.94 kcal.mol$^{-1}$. This whole ORR is carried out with the four-electron transport mechanism. The TS5 and TS6 reaction barriers are important to examine the electrocatalytic efficiency of the 2D monolayer Pt-MoSe$_2$ for ORR. The lower values of the two energy barriers in the ORR show that the 2D monolayer Pt-MoSe$_2$ can be an effective electrocatalyst of ORR in fuel cells.

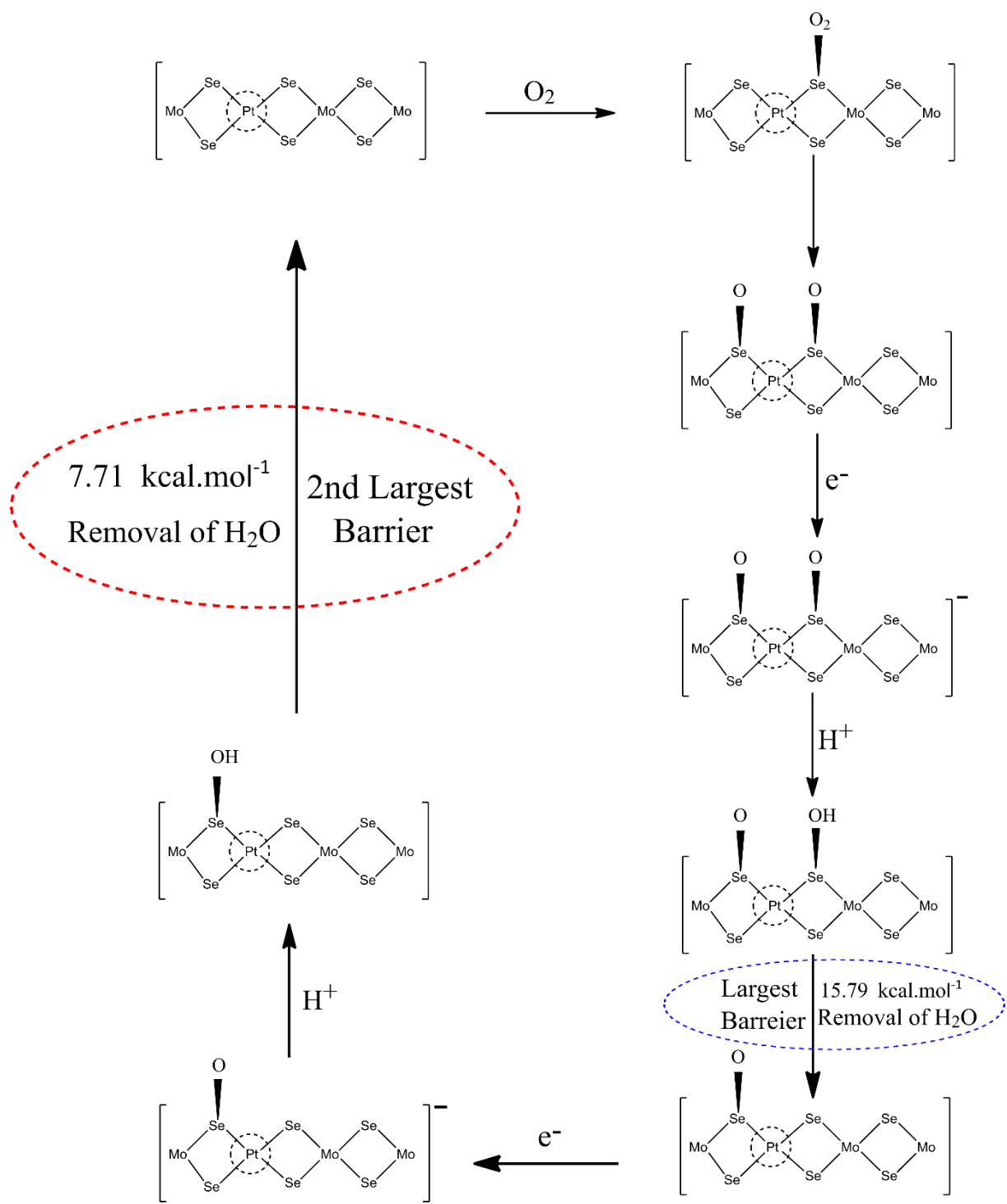

**Fig. 11.** ORR mechanism with the reaction pathway occurring on the surfaces of platinum doped 2D monolayer MoSe$_2$ as an electrocatalyst is shown here and the removal of H$_2$O molecule indicates first TS (i.e., TS5) highlighted in blue ellipse with a barrier energy about 15.79 kcal.mol$^{-1}$ and removal of 2$^{nd}$ H$_2$O gives the second TS (i.e., TS6) highlighted by red ellipse with a barrier of 7.71 kcal.mol$^{-1}$.

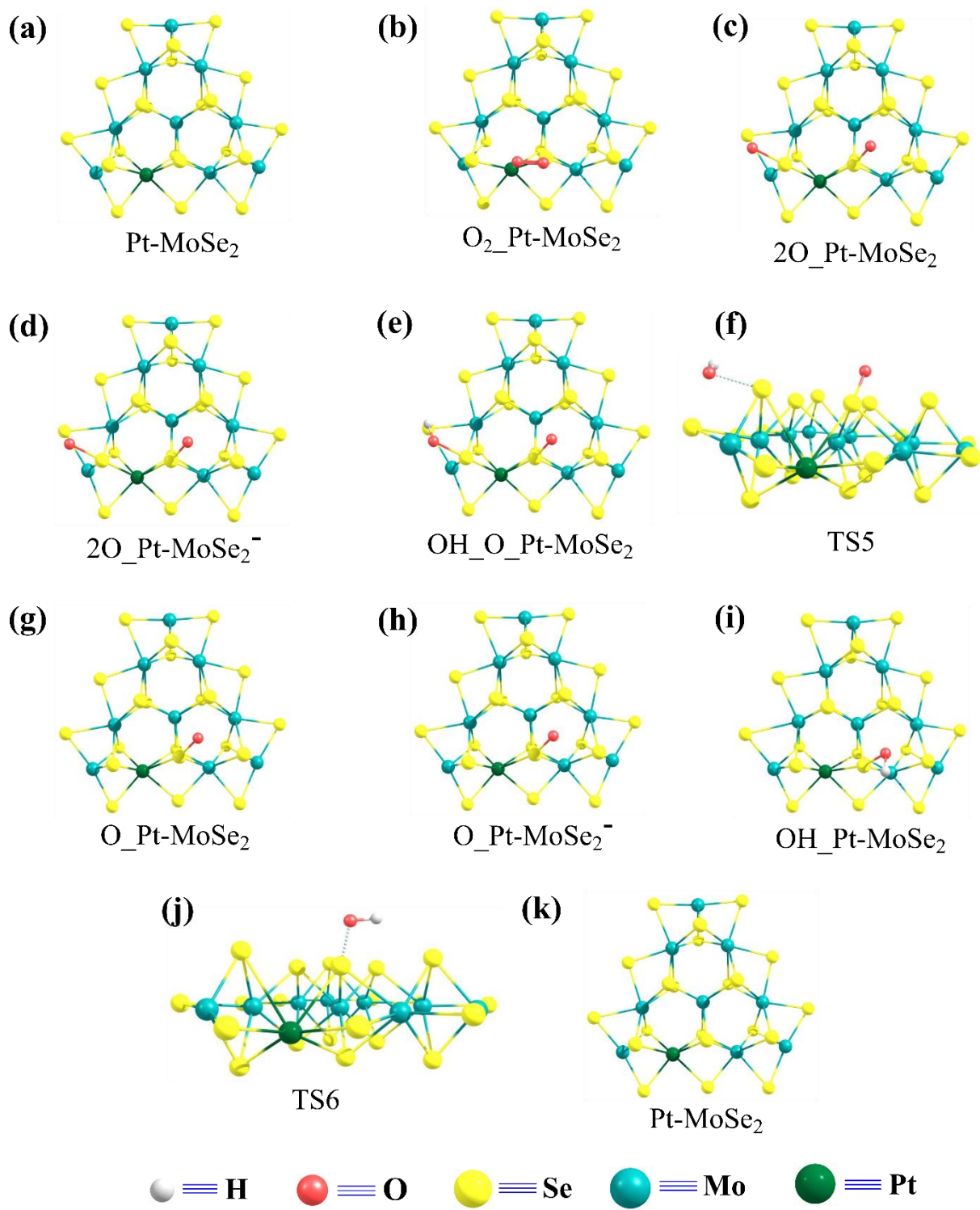

**Fig. 12.** Equilibrium geometries of (a) **[Pt-MoSe$_2$]**, (b) **[O$_2$_Pt-MoSe$_2$]**, (c) **[2O_Pt-MoSe$_2$]**, (d) **[2O_Pt-MoSe$_2$]$^-$**, (e) **[OH_O_Pt-MoSe$_2$]**, (f) **TS5**, (g) **[O_Pt-MoSe$_2$]**, (h) **[O_Pt-MoSe$_2$]$^-$** (i) **[OH_Pt-MoSe$_2$]**, (j) **TS6** (k) **[Pt-MoSe$_2$]** computed by the M06-L DFT method considering a molecular cluster model system Pt$_1$-Mo$_9$Se$_{21}$ to represent 2D monolayer of platinum doped MoSe$_2$ are shown here.

**Table 5.** The changes of various energies of all the intermediates during the ORR performed on the 2D monolayer of Pt- doped MoSe$_2$ (i.e., Pt-MoSe$_2$) are summarized here.

|  | **ORR Reaction Steps** | ΔE kcal.mol$^{-1}$ | ΔG kcal.mol$^{-1}$ | ΔH kcal.mol$^{-1}$ |
|---|---|---|---|---|
| Step I | [Pt-MoSe$_2$] + O$_2$ → [O$_2$_Pt-MoSe$_2$] | -4.32 | 2.74 | -3.37 |
| Step II | [O$_2$_Pt-MoSe$_2$] → [2O_Pt-MoSe$_2$] | -11.38 | -8.11 | -12.17 |
| Step III | [2O_Pt-MoSe$_2$] + e$^-$ → [2O_Pt-MoSe$_2$]$^-$ | -8.14 | -9.79 | -8.27 |
| Step IV | [2O_Pt-MoSe$_2$]$^-$ + H$^+$ → [OH_O_Pt-MoSe$_2$] | -13.38 | -5.36 | -6.09 |
| Step V | [OH_O_Pt-MoSe$_2$] → TS5 | 15.80 | 15.79 | 14.94 |
| Step VI | TS5 → [O_Pt-MoSe$_2$] | -73.68 | -63.69 | -80.43 |
| Step VII | [O_Pt-MoSe$_2$] + e$^-$ → [O_Pt-MoSe$_2$]$^-$ | -27.66 | -28.46 | -27.77 |
| Step VIII | [O_Pt-MoSe$_2$]$^-$ + H$^+$ → [OH_Pt-MoSe$_2$] | -33.58 | -26.15 | -26.15 |
| Step IX | [OH_Pt-MoSe$_2$] → TS6 | 8.36 | 7.71 | 8.00 |
| Step X | TS6 → Pt-MoSe$_2$ | -68.74 | -58.94 | -75.66 |

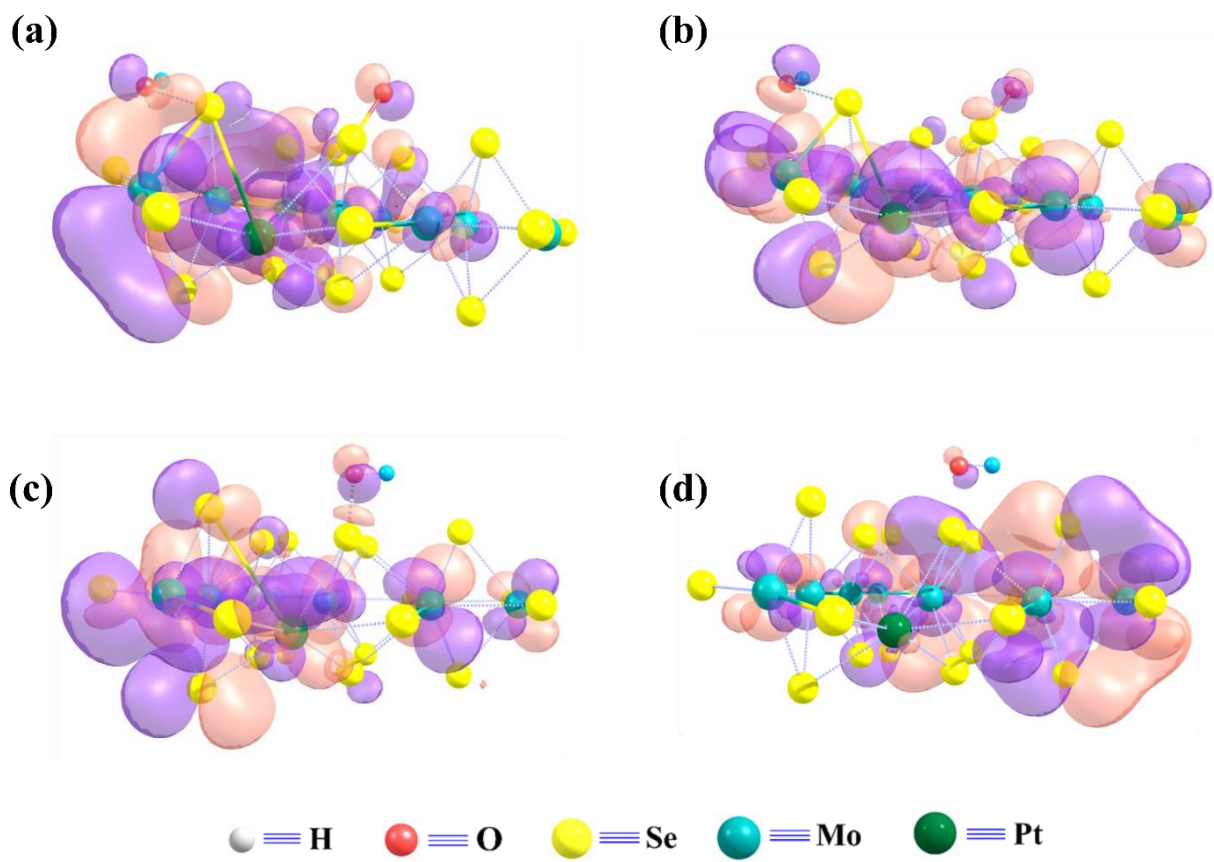

**Fig. 13.** (a) HOMO, (b) LUMO of the transition state structure TS5 and (c) HOMO, (d) LUMO of the transition state structure TS6 occurred in the ORR reaction path are shown here in the case of the 2D monolayer Pt-MoSe$_2$.

**HOMO-LUMO calculation for ORR**

To explain the electrocatalytic ORR activities of the 2D Pt-MoSe$_2$ material, HOMO-LUMO and Frontier molecular orbital (FMO) energies have been computed at the equilibrium geometries of the transition states TS5 and TS6 appeared during the reaction as depicted in Fig. 13a-d. The HOMO – LUMO are central to the FMO theory which resides on the idea that most of chemical reactivity is dominated by the interaction between these orbitals in an electron donor-acceptor pair, in which the most readily available electrons of the former arise from the HOMO and will land at the LUMO in the latter. This theory intuitively assumed that molecular orbitals contributing to chemical reactions were the HOMOs and lowest unoccupied molecular orbitals (LUMOs), which were collectively called "frontier orbitals". According to FMO, the reactive nature of the 2D Pt-MoSe$_2$ surface for oxygen reduction is closely linked to the frontier molecular orbital energies including HOMO and LUMO energies. Increasing the values of HOMO energy show the propensity of a molecule to donate electrons to appropriate electron acceptor while lower LUMO energy exhibits the electrophilic nature of a molecule. Therefore, the lower the LUMO energy, more likely that the molecule will accept the electrons.[67,68] In the cases of TS5 and TS6, the HOMO energy of the 2D Pt-MoSe$_2$ and LUMO energy of H$_2$O were calculated to be -5.64 eV and -0.43 eV, respectively. Here, LUMO energy is relatively higher, and due to this higher value of the LUMO, the electron transfer cannot be possible from catalyst surface to the H$_2$O molecule and hence it would leave the surface instead of getting adsorbed. The atomic orbitals overlapped due to the Pt-doping in the pristine 2D MoSe$_2$ play a significant role in ORR as depicted in Fig. 13. A better overlap of the *p*-orbitals of the oxygen atom and *d*-orbitals of the Pt atom in the 2D Pt-MoSe$_2$ seemed in the HOMO-LUMO transition states TS5 and TS6 during ORR has found, and this better overlap of the atomic orbitals during ORR process reduces the O$_2$ reduction reaction barrier. The better stabilization of the atomic orbitals in the O$_2$ reduction reaction rate-limiting step is a key for reducing the reaction barrier, thus the overall catalysis indicating a better electrocatalytic performance for ORR.

## 4. Conclusions

In the present study, we have computationally designed 2D monolayer pristine MoSe$_2$ and Pt-MoSe$_2$ materials and investigated their electronic properties with electrocatalytic activity by employing first-principles based DFT methods. It has been found that the 2D pristine monolayer MoSe$_2$ shows semiconducting nature with 2.21eV energy bandgap. The inert basal planes of the pristine MoSe$_2$ have been activated by introducing the Pt atoms in the system. After doping the Pt atoms in the pristine MoSe$_2$, the electronic bandgap reduces to 0 eV with a high electron density at the Fermi energy level. This result suggests that Pt-doped MoSe$_2$ material can be used as an efficient electrocatalyst for both the HER and ORR applications. We computationally performed a comparative study of HER using finite non-periodic cluster models of the Mo$_{10}$Se$_{21}$ and Pt$_1$-Mo$_9$Se$_{21}$ followed by Volmer and Heyrovsky mechanism for molecular hydrogen production through electrochemical water splitting. The calculated change in Gibbs free energy for each reaction steps are favorable for the reactions to be spontaneous and thermodynamically stable. The energy barriers of Vomer and Heyrovsky transition structures were found to be 9.29 kcal.mol$^{-1}$ and 10.58 kcal.mol$^{-1}$ in gas phase and 10.55 kcal.mol$^{-1}$ and 11.10 kcal.mol$^{-1}$, respectively, when the HER process takes place on the surfaces of 2D Pt$_1$-Mo$_9$Se$_{21}$. These two reaction barriers of the 2D Pt$_1$-Mo$_9$Se$_{21}$ in both gas and solvents phases are lower than the calculated values of the same two barriers of the 2D pristine $Mo_{10}Se_{21}$ (Volmer and Heyrovsky barriers are 13.44 kcal.mol$^{-1}$ and 20.01 kcal mol$^{-1}$ in the gas phase, and 14.45 kcal.mol$^{-1}$ and 21.75 kcal.mol$^{-1}$ in solvent phase calculations, respectively) with a high TOF. This result suggests that the Pt- doping in the pristine material has enhanced the electrocatalytic activity towards the HER.

An investigation on the $O_2$ reduction reaction has been performed on the surfaces of 2D Pt-$MoSe_2$ considering the same molecular cluster $Pt_1$-$Mo_9Se_{21}$ model system. The complete reaction mechanism with the pathway has been explored by computing the free energy changes each step of the ORR, and all the reaction steps were found to be spontaneous and thermodynamically favorable. The ORR on the surface of 2D monolayer Pt-$MoSe_2$ material has been achieved with a four-electron transport mechanism including the formation of two transition states TS5 and TS6 with the energy barriers about 15.79 kcal.mol$^{-1}$ and 7.71 kcal.mol$^{-1}$, respectively. These two TSs have been formed when the $H_2O$ was removed from the surface and these structures were confirmed for being TSs from the observation of negative frequency value, modes of vibrations and IRC calculations. Both the TSs (i.e., TS5 and TS6) during the ORR have a very low energy barriers, which indicates that the 2D Pt-$MoSe_2$ has an efficiency for the electrocatalytic applications for ORR same as HER. HOMO and LUMO calculations have been performed to understand the reason behind the high catalytic activities of the Pt-doped $MoSe_2$ material for both the HER and ORR. In the HER process, the Heyrovsky reaction step i.e., TS4 has been formed due to the overlapping of the atomic orbitals of two H atoms with the adjacent hydronium ($H_3O^+$) in the explicit water cluster to form the $H_2$ molecules during HER. These overlapping with well stabilization (compared to the pristine 2D monolayer $MoSe_2$) determines the lower reaction barrier and electrocatalytic performance of the 2D Pt-$MoSe_2$ TMD for effective HER. Similarly, a superior stabilization of the atomic orbitals in the $O_2$ reduction reaction (i.e., rate-limiting step) is a key for reducing the reaction barrier, thus the overall catalysis indicating a better electrocatalytic performance for ORR. In summary, we can lead our current discussion for a conclusion that the computationally developed 2D monolayer Pt-$MoSe_2$ material can be used as an effective electrocatalyst for both the ORR and HER applications as a bifunctional electrocatalyst.

## AUTHOR CONTRIBUTION

Dr. Pakhira developed the complete idea of this current research work with project and Mr. Shrish Nath Upadhyay computationally studied the electronic structures and properties of the 2D monolayer Pt-MoSe$_2$. Dr. Pakhira and Mr. Upadhyay explored the whole reaction pathways; transitions states and reactions barriers and they explained both the ORR and HER mechanism by the DFT calculations. Quantum calculations and theoretical models were designed and performed by Dr Pakhira and Mr. Upadhyay. Dr. Pakhira and Mr. Shrish Nath Upadhyay wrote the manuscript and prepared all the tables and figures in the manuscript.

**Notes**

The authors declare no competing financial interest.

## Supplementary Information:

This work was carried out with the following version of the programs. All the structures were modeled by using the VESTA software. (https://jp-minerals.org/vesta/en/) All the calculations were carried out by using the *ab-initio* based CRYSTAL17 suite code. (https://www.crystal.unito.it/) GNUPLOT and Inkscape 0.92 were used to plot the electronic band structures, total DOS and PES. (http://www.gnuplot.info/ and https://inkscape.org/). Python scripts were used to generate data plots. (https://www.python.org/) The Supplementary Information is available free of charge on the RSC Publications website. All the equilibrium structures involved in the subject reaction have been provided in the Supplementary Information. All the equilibrium structures with their crystallographic information files (.cif) involved in the subject reaction have been provided in the Supplementary Information available in the website: https://www.rsc.org/journals-books-databases/about-journals/pccp/.

## AUTHOR INFORMATION

**Corresponding Author**
**Dr. Srimanta Pakhira** − *Department of Physics, Indian Institute of Technology Indore (IITI), Khandwa Road, Simrol, Indore, MP 453552, India.*

*Discipline of Metallurgy Engineering and Materials Science, Indian Institute of Technology Indore (IITI), Khandwa Road, Simrol, Indore, MP 453552, India.*
*Centre of Advanced Electronics (CAE), Indian Institute of Technology Indore, Khandwa Road, Simrol, Indore, MP 453552, India.*
ORCID: orcid.org/0000-0002-2488-300X.
Email: spakhira@iiti.ac.in and spakhirafsu@gmail.com

**Author**

**Shrish Nath Upadhyay** − *Discipline of Metallurgy Engineering and Materials Science (MEMS), Indian Institute of Technology Indore (IITI), Khandwa Road, Simrol, Indore, MP 453552, India*.
ORCID: orcid.org/0000-0003-0029-4160.

## ACKNOWLEDGEMENTS

This work was financially supported by the Science and Engineering Research Board-Department of Science and Technology (SERB-DST), Government of India under Grant No. ECR/2018/000255. Dr. Srimanta Pakhira acknowledges the SERB-DST, Government of India for providing his Early Career Research Award (ECRA) under the project number ECR/2018/000255, and highly prestigious Ramanujan Faculty Fellowship under the scheme number SB/S2/RJN-067/2017. Funding of this research in Dr. Pakhira's research group at IIT Indore is supported by the SERB-DST, Government of India. We thank the SERB-DST, Govt. of India for financial assistance by providing the Core Research Grant (CRG) under the scheme number CRG/2021/000572. Mr. Upadhyay thanks Indian Institute of Technology Indore, MHRD, Govt. of India for providing the doctoral fellowship. The support and the resources provided by PARAM Brahma Facility under the National Supercomputing Mission (NSM), Government of India at the Indian Institute of Science Education and Research, Pune, India are gratefully acknowledged. We thank the respected reviewers for providing their valuable comments and suggestions to improve the quality of the manuscript.

## Conflicts of Interest:

The authors have no additional conflicts of interest.

**TOC: Graphical Abstract**

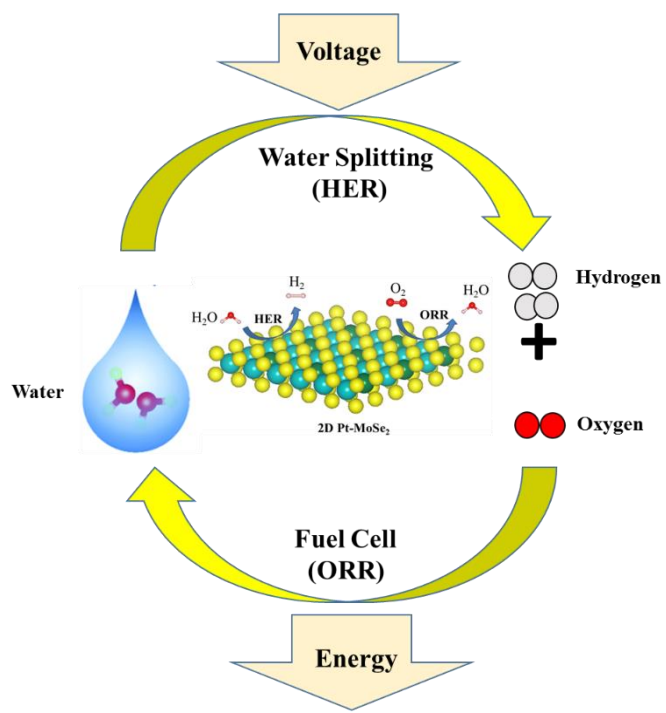

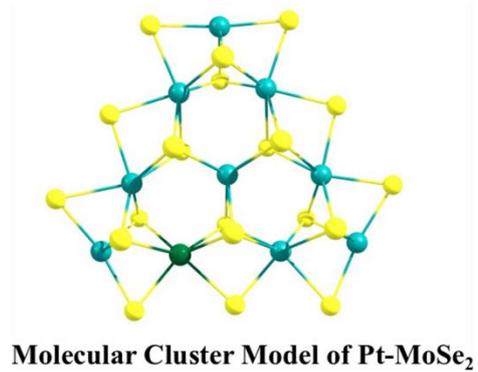

Molecular Cluster Model of Pt-MoSe$_2$

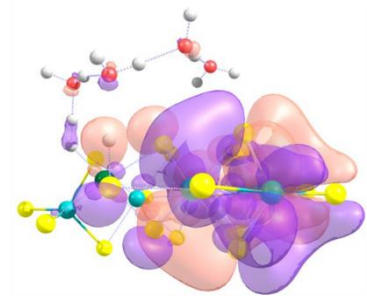

HOMO of TS4 in Heyrovsky Step